\documentclass[amsmath,amssymb,amsbsy,aps,prx,twocolumn,superscriptaddress,longbibliography,floatfix]{revtex4-2}
\usepackage{amsfonts} 
\usepackage{amsmath} 
\usepackage{amssymb}
\usepackage{cancel}
\usepackage{bm}
\usepackage[usenames, dvipsnames]{color} 
\usepackage{dcolumn}
\usepackage{epic} 
\usepackage{epsfig}
\usepackage{esdiff}
\usepackage{graphicx}
\usepackage[breaklinks,colorlinks = true,linkcolor = magenta,urlcolor=magenta,citecolor=red]{hyperref}
\usepackage{mathrsfs}
\usepackage{soul}
\usepackage{subfigure}
\usepackage{textcomp}
\usepackage{times}
\usepackage{verbatim}
\usepackage{wrapfig} 
\usepackage{xy} 

\newcommand{\commentout}[1]{}

\newcommand{\vect}[1]{\mathbf{#1}}

\newcommand{\JILA}{JILA, National Institute of Standards and Technology and Department of Physics, University of Colorado, Boulder, CO, 80309, USA}
\newcommand{\CTQM}{Center for Theory of Quantum Matter, University of Colorado, Boulder, CO, 80309, USA}
\newcommand{\MPIPKS}{Max-Planck-Institut f\"{u}r Physik komplexer Systeme, N\"{o}thnitzer Strasse 38, 01187 Dresden, Germany}

\begin{document}
\title{Classical many-body chaos with and without quasiparticles}

\author{Thomas Bilitewski}
\email{thomas.bilitewski@colorado.edu}
\affiliation{\JILA}
\affiliation{\CTQM}
\affiliation{\MPIPKS}
\author{Subhro Bhattacharjee}
\email{subhro@icts.res.in}
\affiliation{International Centre for Theoretical Sciences, Tata Institute of Fundamental Research, Bengaluru 560089, India}
\author{Roderich Moessner}
\email{moessner@pks.mpg.de}
\affiliation{\MPIPKS}
\date{\today}
\begin{abstract}
We study correlations, {transport} and chaos in a Heisenberg magnet as a classical model many-body system. By varying temperature and dimensionality, we can tune between settings with and without symmetry breaking and accompanying collective modes or 
quasiparticles (spin-waves) which in the limit of low temperatures become increasingly long-lived. 
Changing the sign of the exchange 
interaction from ferro- to an antiferromagnetic one varies the spin-wave spectrum, and hence the low-energy spectral density. We analyse both conventional and 
out-of-time-ordered spin correlators  (`decorrelators') to track the spreading of a spatiotemporally localised perturbation -- the wingbeat of the butterfly --  as well as {transport coefficients 
and} Lyapunov exponents. We identify a number of qualitatively different regimes. Trivially, at $T=0$, there is no dynamics at all. In the limit of low temperature, $T=0^+$, integrability emerges, with infinitely long-lived  magnons; here the wavepacket created by the perturbation propagates ballistically, yielding  a lightcone at the spin wave velocity which thus subsumes the butterfly velocity; inside the lightcone, a pattern characteristic of the free spin wave spectrum is visible at short times. On top of this, residual interactions (nonlinearities in the equations of motion) lead to spin wave lifetimes which, while divergent in this limit, remain finite at any nonzero $T$. At the longest times, this leads to a `standard' chaotic regime; for this  regime, we show that  the Lyapunov exponent is simply proportional to  the  (inverse)  spin-wave lifetime. Visibly strikingly, between this and the `short-time' integrable regimes, a scarred regime emerges: here, the decorrelator is spatiotemporally highly non-uniform, being dominated by rare and random scattering events seeding secondary lightcones. As the spin correlation length decreases with increasing $T$, the distinction between these regimes disappears and at high temperature  the previously studied chaotic paramagnetic regime emerges. For this,  we  elucidate how, somewhat counterintuitively, the {\it `ballistic'} butterfly velocity arises from a {\it diffusive} spin dynamics.
\end{abstract}

\maketitle
\section{Introduction}
The study of many-body chaos has gained new momentum at the confluence of two developments. One is the ability to access coherent quantum many-body dynamics in a variety of experimental platforms \cite{Browaeys2020,Blatt2012,Bloch2012,Houck2012,Aspuru-Guzik2012,PhysRevX.7.041063}. The other is the nature of information scrambling in quantum field theories of strongly correlated condensed matter and gravity including their possible interconnections.
\cite{Maldacena_1999,maldacena2016bound,Sekino_2008,PhysRevD.78.046003,PhysRevLett.101.061601,Roberts2015,Shenker2014b,Cotler2017,Stanford2016,kitaev,gu2017local}. 

A central quantitative measure of (quantum) chaos that has emerged in this recent progress  are out-of-time ordered commutators (OTOC) \cite{1969LarkinOvchin} which, in a class of quantum many-body systems, show exponential temporal growth and or ballistic propagation akin to classical chaos also in quantum systems \cite{PhysRevLett.115.131603,PhysRevLett.119.026802,PhysRevLett.118.086801,PhysRevB.96.020406,PhysRevB.96.060301,ALEINER2016,swingle2016measuring,swingle2017slow,PhysRevB.97.144304,Chen2017,nahum2017operator,von2017operator,PhysRevX.8.031058,PhysRevX.8.031057,Xu_2019}. Broadly speaking, these are found to either have a sharp front propagating ballistically, e.g. in large N and coupled Sachdev-Ye-Kitaev (SYK) models \cite{Xu_2019,gu2017local,Chowdhury_2017}, or show diffusively broadening fronts as in random circuit models with and without conservation laws \cite{brown2012scrambling,nahum2017operator,von2017operator,PhysRevX.8.031057,PhysRevX.8.031058}, which is universally captured by velocity-dependent Lyapunov exponents \cite{khemani2018velocity,Xu_2019}. 

In parallel the classical versions of OTOCs (alternatively dubbed as {\it decorrelators}) have been applied to study the spatiotemporal chaos in classical many-body systems  \cite{PhysRevLett.121.024101,PhysRevLett.121.250602,kumar2019many,Schuckert_2019,ruidas2020manybody}, in part with the goal to elucidate their significance in the more conventional realm of classically chaotic models, but also to understand the semi-classical limit of many-body   quantum chaos. In case of chaotic classical many-body systems with short range correlations, the study of the decorrelators clearly reveals two complementary aspects of the {\it butterfly effect}-- (1) the exponential temporal growth of a localised (in real space) infinitesimal difference in the initial conditions characterised by the Lyapunov exponent, $\lambda$, and (2) its ballistic spread  characterised by the {\it butterfly speed}, $v_B$ \cite{PhysRevLett.121.024101,PhysRevLett.121.250602}.

A natural question then pertains to the dependence and relation between the above chaos time-scales, $\lambda^{-1}$, and length-scales, $v_B\lambda^{-1}$, on the thermodynamic and dynamic properties of the system. This assumes particular importance with regard to two  central results about chaos in quantum-many body systems where it has been shown that: (1) in maximally chaotic system, the Lyapunov exponent is universally bounded by the absolute temperature as $\lambda_{\rm max}=2\pi k_B T/\hbar$,\cite{maldacena2016bound} and (2) under selected circumstances, the diffusion constant, $D$, of a conserved charge and the chaos time and length scales are related via $D\sim v_B^2/\lambda$.

The above two questions remain equally important in the context of classical many-body systems, where it has recently been shown that the maximal bound is generically `violated' in the disordered phase of spin-rotation symmetric classical spin systems and thermalised fluids where $\lambda\propto \sqrt{T}$ is observed \cite{PhysRevLett.121.024101,PhysRevLett.121.250602,kumar2019many,ruidas2020manybody}: the limits of temperature $T\rightarrow0$ and spin $S\rightarrow \infty$ do not commute. In particular, in a classical spin liquid that remains disordered down to $T=0$, the above behaviour was found down to the lowest temperatures as well as $D_S\sim v_B^2/\lambda$ for the spin diffusion \cite{PhysRevLett.121.250602}. 

The interplay of symmetries and finite temperature however is much  richer in many-body systems including the possibility of spontaneous symmetry breaking at low temperatures through a thermal phase transition that is accessed by lowering the temperature. 

In this paper, we concern ourselves with the study of the spatiotemporal chaos in a paradigmatic many-body system of classical spins that is tuned through a spontaneous symmetry breaking magnetic phase transition at finite temperatures, or that at least exhibits a divergent correlation length at low temperatures. In particular, we consider short range interacting spin rotation invariant Heisenberg Hamiltonians for classical spins  sitting on a $d$-dimensional hypercubic lattice (with $d=1,2,3$):
\begin{align}
H=J \sum_{\langle ij\rangle} {\bf S}_{i}\cdot{\bf S}_j
\label{eq_ham}
\end{align}
where ${\bf S}_i$ are unit vectors encoding classical three component spins at each lattice site, $i$, $J$ is the nearest-neighbour exchange constants for spins joined by bond $\langle ij\rangle$, and we study both ferromagnetic, $J<0$, and antiferromagnetic, $J>0$, cases. The dynamics we consider is precessional, i.e. given by the Landau-Lifshits equations of motion, Eq.~\ref{eq_eom} \cite{Landau_1999,spin_dyn_sim}, which are generated either by canonical Poisson brackets in the classical case, or commutators for quantum systems. 

Indeed, an  advantage of this model system lies in such a direct connection of the classical model to the quantum system in the form of a semi-classical limit, where observables tracking the chaotic properties can easily be constructed to apply to both cases \cite{PhysRevLett.121.024101} providing a potentially more direct connection of classical and quantum chaos than in other situations.

We find that in practice, even though the Heisenberg magnet in  $d=1$ has no true long-range order at any nonzero temperature \cite{PhysRevLett.17.1133}, that it is most convenient to focus our numerical efforts on this case: thanks to its divergent correlation length in the limit of zero temperature, it does exhibit a regime
where for practical purposes, long-range order and a finite spin-wave velocity exist~\cite{tomita1972spin}, while large linear system sizes, up to $L=4\times 10^4$ spins, are numerically tractable.
However, we also present numerical data on systems in $d=2,3$ with a similar number of spins but correspondingly smaller linear sizes. Indeed, most of our results are generally consistent across dimensions and in absence ($d=1,2$) or presence ($d=3$) of a true thermal phase transition. The situations where the presence or absence of the phase transition is of essential importance -- as in the case of the temperature dependence of the butterfly velocity (Fig.~\ref{fig:v_vs_T}) -- are discussed at their respective places.

We provide a systematic understanding of the following questions.  Are there distinct regimes for chaos in accordance with these phases, and if so, what are their properties? How can we characterise them through observables? What role do quasiparticles, present in systems with spontaneously broken symmetry, play?  A first step towards answering these questions was the study of the temperature dependence of chaos in absence of phase transitions in a classical spin system \cite{PhysRevLett.121.250602}. 
There are of course numerous precursors to this work which have considered 
spatiotemporal chaos in classical many-body systems, often focussing on 
high/infinite temperatures, or without a notion of temperature at all \cite{PhysRevLett.80.692,PhysRevLett.80.2035,gerling1990time,de2012largest,PhysRevE.70.016207,PhysRevLett.109.034101,de_Wijn_2013}.
More recently, many-body chaos was studied close to thermal phase transitions in a scalar field theory \cite{Schuckert_2019}, and in a classical XXZ spin system in two dimensions \cite{ruidas2020manybody}, where qualitatively different behaviour of the chaos quantities, $v_B$ and $\lambda$, was found at or below/above the phase transition.

Using a combination of direct numerical simulations and mode-coupling calculations for the spin-waves, we reveal the features of the short time {\it emergent integrability} at low temperatures ultimately giving away to chaos at intermediate times paving way for the long-time thermalisation. 

The chaotic behavour is quantitatively characterised via  the decorrelator, the specific measure at the centre of this investigation, which we define in Sec.~\ref{sec:decaff}, Eq.~\ref{eq_decorrelator}. 
This is part of a detailed introduction to the system and the observables to be studied, to which Sec.~\ref{sec:introdef} is devoted. 

Our narrative then proceeds from low to high temperatures, as outlined in the abstract. 
The low-$T$ ordered regime is the subject of Sec.~\ref{sec:lowT}. There, we study in detail the spatio-temporal behaviour of the decorrelator which we use to characterise the many-body chaos. We find that for short times, it is well described by non-interacting spin wave theory. In particular, we find ballistic propagation of an initially localised perturbation, and an initial powerlaw decay of the decorrelator consistent with this ballistic spreading of spin-waves. Remarkably, the butterfly velocity, i.e.\ the speed with which the light-cone advances, continues to be  given by the quasiparticle velocity from linear spin wave theory, even when the decorrelator overall is dominated by the expontential growth characteristic of spatiotemporally chaos.

Furthermore, at low temperatures we find a regime of a ``scarred" decorrelator. This is sandwiched between the short-time integrable  and the late time chaotic regimes. It arises through repeated scatterings of the long-lived, well-defined propagating quasiparticles, which seed `secondary' lightcones. The superposition of many such secondary  lightcones then generates the `diffusive' core of late-time chaos. This diffusive core grows parametrically more slowly than the the (primary) ballistic lightcone, inside which it is located.

In Sec.~\ref{sec:T_dep}, we then present the behaviour of the butterfly velocity, characterising the spatial propagation of chaos, and the Lyapunov exponent, characterising the exponential temporal growth, as a function of temperature in $d=1,2,3$. We find that both characteristically change behaviour at the (finite-size) thermal phase `transition'. Whereas the Lyapunov vanishes as a powerlaw in temperature below the transition,  the velocity 
saturates to a finite value both in the paramagnetic high-temperature and the magnetically ordered low temperature regime. 

\begin{figure*}[!htp]
\includegraphics[scale=1.1]{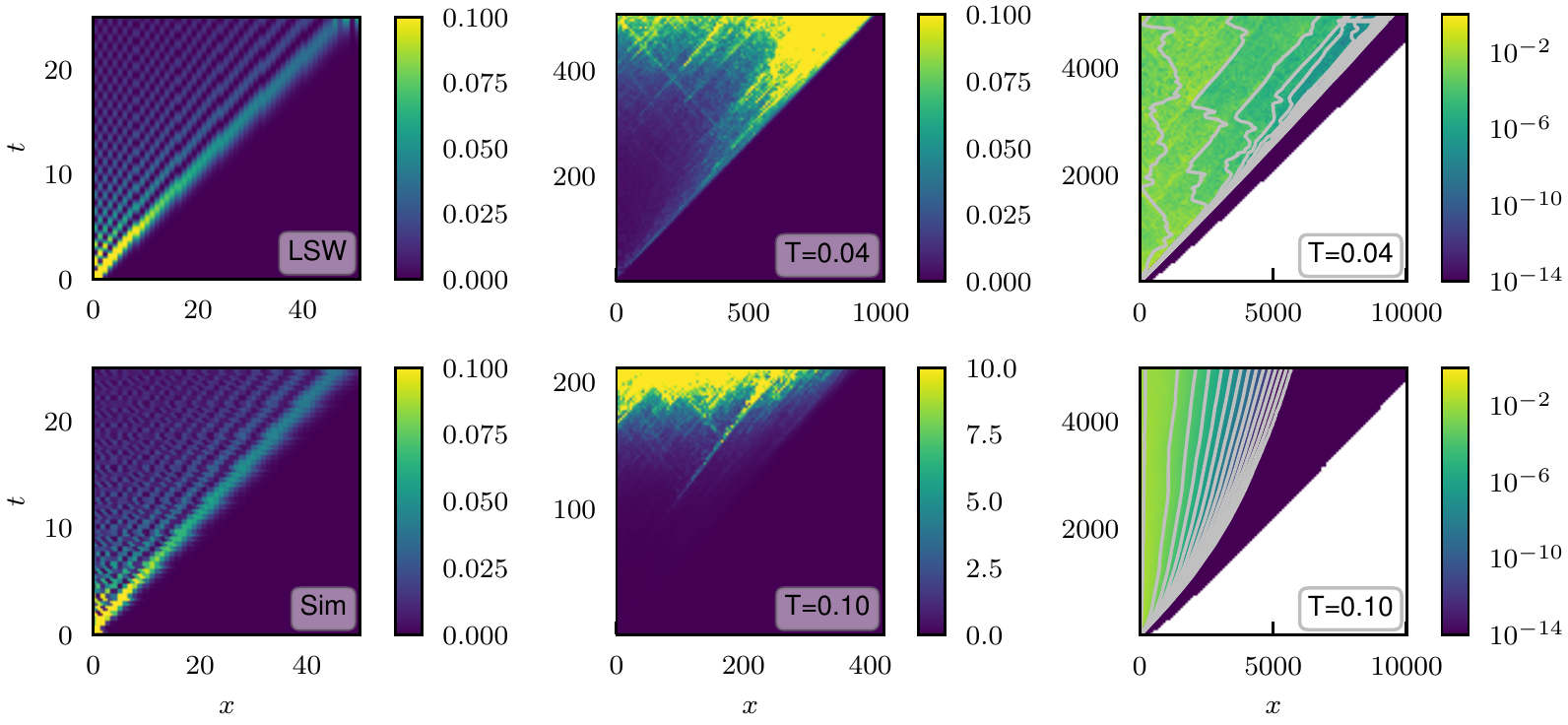}
\caption{Different regimes of the decorrelator $D(x,t)$ for  one dimensional nearest neighbour ferromagnet. Left panels: short time integrable regime comparing the linear spin wave (top) to the full numerical simulations (bottom) at $T=0.04$. Middle panels: intermediate ``scarred" regime at $T=0.04$ (top) and $T=0.10$ (bottom). Right panels: long time regime, where repeated scattering results in diffusive scaling. Here, the scaled decorrelator $D(x,t) = D(x,t)/\int dx D(x,t)$ is shown on a logarithmic colorscale. Gray lines are contours on a logarithmic grid between $10^{-1}$ and $10^{-14}$. }
\label{fig:summary}
\end{figure*}

In Sec.~\ref{sec:SL_vs_FM} we then contrast the behaviour of the ordered magnet to the previously studied case of the spin-liquid, and discuss the high-temperature paramagnetic phase, where diffusive and ballistic behaviour co-exists.
Most notably, we find a direct relation, in the form of a proportionality, between the Lyapunov exponent and a combination of a  characteristic velocity of the quasiparticles, and their associated scattering rates.  From kinetic theory, this leads to the relation $D\sim v_\mathrm{qp}^2\tau_\mathrm{qp} \sim v_B^2/\lambda$: the quasiparticle velocity plays the role of the butterfly velocity.
Finally, for the high-temperature short-range correlated paramagnetic regime, we show how the diffusive nature of the spin transport is consistent with the ballistic spreading of the decorrelator front.
This section also makes contact with our previous study of chaos in a cooperative paramagnet on the kagome lattice \cite{PhysRevLett.121.250602}, which exhibits scaling forms for chaotic observables down to $T=0$, and which is relevant to the high-temperature paramagnetic phase of the hypercubic Heisenberg models studied here.
We conclude with a discussion Sec.~\ref{sec:discussion}.

For the benefit of the reader interested in a qualitative impression of our central results without having to wade through the copious details, Sec.~\ref{sec:sketch} provides a summary of these in the form of what we hope is a visually compelling survey of the behaviour of decorrelators in the various regimes.

\section{Overview of the regimes}
\label{sec:sketch}

Here, we present the different regimes of many-body chaos that we have identified. To unclutter the presentation, we concentrate on the case of the $d=1$ Heisenberg ferromagnet, and present corresponding figures for antiferromagnets, as well as  higher $d$, in later sections. All figures display the spatiotemporal behaviour of the decorrelator, Eq.~\ref{eq_decorrelator}, on different timescales, and at different temperatures. 

{\it Integrable regime:} 
 The left panels of Fig.~\ref{fig:summary} show the short-time spread of the perturbation evaluated in two different ways. The top one displays the results of linear spin-wave theory, a completely non-interacting theory without scattering or chaos. The lower panel displays the result of our numerical simulations. The two plots agree in considerable detail, exhibiting the following features. First, there is a clear lightcone, advancing with velocity $v_B\sim |J|$, beyond which the signal vanishes rapidly. The amplitude decays as a powerlaw $t^{-1}$ within this light-cone. Second, there is considerable structure in the decorrelator, which reflects the properties of the spin waves across the entire spectrum, as the localised initial perturbation excites all of them. 

{\it Scarred regime:} At longer times, rare scattering event become visible, middle panels of Fig.~\ref{fig:summary}. In these, secondary lightcones appear at random points in spacetime, which have ballistically propagating edges  in both directions. As an increasing number of these overlap as time progresses, a distinctive scarred appearance emerges. 

{\it Diffusive regime:} Eventually, the number of secondary lightcones becomes so large that a given point in space is under the influence of many of them, leading to a statistical average. This is determined by the `random walk' of the lightcones, and therefore takes on a standard diffusive shape, with equipotentials of the decorrelator following trajectories $\langle x^2\rangle\sim t$. Note that the crossover between the scarred and the diffusive regime can take place at rather late times: for $T=0.04$ at $t=1000$, scars are still plainly visible even after averaging over $10^3$ initial states.

This diffusive regime grows at the expense of the other two as $T$
 is raised. This connects to the analysis of the spin liquid regime presented in Ref~\cite{PhysRevLett.121.250602}, where this is the only regime present across all temperatures.

\section{Model and observables}
\label{sec:introdef}
In this section, we first introduce the central object of study for much of this work, namely the decorrelator. We then describe the thermal properties of our model system, followed by setting the stage for the analysis of its dynamics.

\subsection{The decorrelator}
\label{sec:decaff}
We start by introducing the decorreletor: It measures the divergence of two copies in response to a perturbation applied to one of them, which can be chosen to be spatiotemporally localised, as in the wingbeat of the butterfly. This is an instance of an out of time ordered correlator, and was first introduced in Ref.~\cite{PhysRevLett.121.024101} to study the classical many-body chaos in a spin chain at infinite temperature. 

Specifically, we label the two copies of the same spin system I and II, which both evolve under the same Hamiltonian through the equation of motion given by Eq.~\ref{eq_eom}. We  consider an infinitesimal difference in the initial condition (i.e., at time $t=0$) of the two copies only at site $i=0$:
\begin{align}
    \delta{\bf S}_i={\bf S}_i^{\mathrm{II}}-{\bf S}_i^{\mathrm{I}}=\boldsymbol{\epsilon}\delta_{i,0}=\varepsilon({\bf n}\times {\bf S}_{i,0})\delta_{i,0}
    \label{eq_dels}
\end{align}
where $0<\varepsilon\ll 1$, and
\begin{align}
    \hat{\bf n}=\frac{\hat{\bf z}\times{\bf S}_0(0)}{|\hat{\bf z}\times{\bf S}_0(0)|}\ .
    \label{eq_delsn}
\end{align}
The decorrelator $\mathcal{D}({i},t)$  \cite{1969LarkinOvchin,PhysRevE.89.012923,PhysRevLett.121.024101,PhysRevLett.121.250602,kumar2019many}, defined as the local difference of the spin configuration of the two copies after a time $t$ 
\begin{align}
\mathcal{D}({i},t)=\left<\left({\bf S}_{{i},t}^{\mathrm{II}}-{\bf S}_{{i},t}^{\mathrm{I}}\right)^2\right>_T=\left<\vert\delta{\bf S}_{i,t}\vert^2\right>_T \, ,
\label{eq_decorrelator}
\end{align}
(where $\langle\cdots\rangle_T$ denotes averaging over a set of initial conditions chosen from a canonical ensemble of spin configurations at temperature $T$), measures the temporal growth and spatial spread of the difference in initial condition. In this sense $\delta{\bf S}_{i,t}$ is the {\it difference field} that evolves spatially outward from $i=0$ with time. 

It will be useful to introduce the Fourier modes for the various low energy fields. In particular,
\begin{align}
\delta{\bf S}_i=\frac{1}{N}\sum_{\bf k}e^{-i{\bf k\cdot r_i}}\delta{\bf S}_{\bf k}
\label{eq_fts}
\end{align}
where $\delta{\bf S}_{\bf k}$ is the Fourier mode with with momentum ${\bf k}$ such that the initial condition (Eq.~\ref{eq_dels}) becomes
\begin{align}
\delta{\bf S}_{\bf k}(t=0)=\boldsymbol{\epsilon}
\label{eq_sk_flat}
\end{align}

The decorrelator in Eq.~\ref{eq_decorrelator} is now given by
\begin{align}
\mathcal{D}(i,t)&=\frac{1}{N^2}\sum_{\bf kk'}e^{-i({\bf k}+{\bf k'})\cdot{\bf r}_i}~\langle\delta {\bf S}_{\bf k}(t)\cdot\delta{\bf S}_{\bf k'}(t)\rangle_T
\label{eq_decorexp}
\end{align}
which contains the entire spatiotemporal information. We shall also find it useful to introduce the space averaged decorrelator defined by
\begin{align}
    \mathcal{I}(t)=\frac{1}{N}\sum_{i}\mathcal{D}(i,t)=\sum_{\bf k} ~\langle\delta {\bf S}_{\bf k}(t)\cdot\delta{\bf S}_{-\bf k}(t)\rangle_T \, ,
    \label{eq_sumdecorexp}
\end{align}
 which, due to the averaging over the spatial information, such as the beating patterns discussed above, conveniently allows to isolate the temporal evolution.
 This space-averaged de-correlator is similar to the {\it distance function} studied in Ref. \cite{derrida1986phase} in the context of cellular automata. 

We note that the question of the dynamics of a single misaligned spin in a ferromagnetically ordered background was investigated several years ago in Refs. \cite{PhysRev.181.811} and \cite{PhysRev.146.387}. While these studies did find ballistic spread in the magnetically ordered state, the high-temperature phase was assumed to be purely diffusive. However, as shown in Ref. \cite{PhysRevLett.121.024101} the decorrelator displays ballstic spreading even at infinite temperature.

\subsection{The classical Heisenberg magnet}
The model we study is the classical Heisenberg model, the  Hamiltonian of which is given by Eq.~\ref{eq_ham}. In this work we shall exclusively consider nearest neighbour interactions on a set of bipartite hypercubic lattices with varying dimensionality: (1) one dimensional chain, (2) two dimensional square lattice, and (3) three dimensional cubic lattice. We consider both ferromagnetic and antiferromagnetic interactions. 

As the system is spin rotation invariant the three components of the total magnetisation
\begin{align}
{\bf S}_T=\sum_{i} {\bf S}_{i}
\label{eq_stotal}
\end{align} 
are conserved along with the total energy $E=J \sum_{ij}  \vect{S}_i \cdot \vect{S}_j$. The fate of these conservation laws influence the thermodynamic and transport properties of the system across the entire temperature regime.

We also note that due to the conservation of both energy and magnetisation during the dynamics,  generically any observable computed from dynamical trajectories, such as the decorrelator described above, is a function of the total magnetisation $\vect{S}_T$ and energy $E$ (and any other conserved quantity). A given ensemble, e.g. a thermal ensemble of initial states at temperature $T$, then corresponds to averages over the corresponding (thermal) distributions of the conserved quantities. However a micro-canonical ensemble at fixed $\vect{S}_T$ and $E$ contains equally valid information about the dynamical properties of the model. In particular, this implies that a finite-size system with a finite magnetisation (even if the thermodynamic limit would show a vanishing magnetisation) can shed light on the dynamics in a symmetry-broken state.

\subsubsection{The thermodynamics: phases and phase transitions}
Our central tuning parameter is temperature. This allows us to access regimes with short-range correlations, and (at least effectively) long-range order. 

\begin{figure}
    \centering
    \includegraphics[scale=0.95]{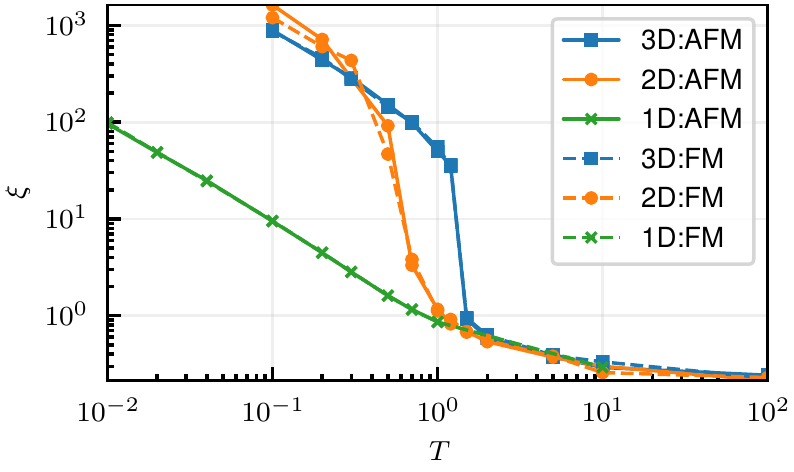}
    \includegraphics[scale=0.95]{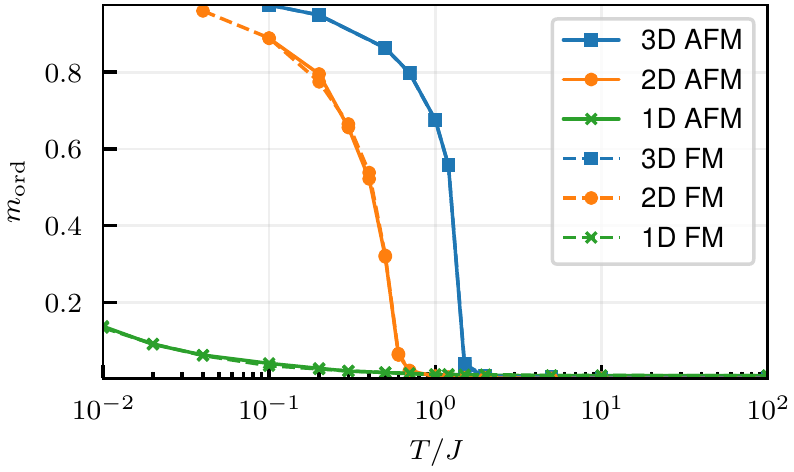}
    \caption{Finite-size correlations. Top: Spin correlation length $\xi$ versus temperature $T$ defined via fits to $\left< S_{\vect{r}} S_{\vect{r}'} \right>= e^{- \left|\vect{r}-\vect{r}' \right|/\xi}$. Note that below the transition correlations are not exponentially decaying on finite systems, and $\xi$ rather qualitatively tracks the long-ranged nature. this Bottom: Order parameter $m_{ord} = (1/L^d) \sum_i S_i$ or $(1/L^d)\sum_i (-1)^i S_i$ for the FM and AFM respectively versus temperature $T$. Both for the FM (dashed) and AFM (solid) for systems in 1D ('crosses') with $L=10000$, 2D ('circles') with $L=400$ and 3D ('squares') with $L=40$. 
    \label{fig:xi_vs_T}}
\end{figure}

At high temperatures, a spin rotation invariant thermal paramagnet is always realised, where the spin correlations are short ranged. 
At low temperature in $d=3$, there is a magnetic ordering transition at a  critical temperature, $T_c$ \cite{PhysRevB.48.3249}. By contrast, for $d\leq2$ a thermal phase transition at a nonzero temperature is absent \cite{PhysRevLett.17.1133}. Instead, the correlation length diverges in the limit $T\rightarrow 0$ in $d=1,2$. As a result, at sufficiently low temperatures, the correlation length will always exceed the size of any finite lattice. In such a regime, 
the behaviour of the system in many respects resembles that of a symmetry-broken phase-- {\it i.e.}, a uniform magnetised phase for the ferromagnetic interaction and a Neel order phase for the antiferromagnetic interactions. 

We illustrate this behaviour  in Fig.~\ref{fig:xi_vs_T} for both the ferromagnet and the antiferromagnet model in $d=1,2,3$ by considering the behaviour of the spin correlation length $\xi$ versus temperature $T$. 
Note that, whereas both two and three dimensional systems show a rather well-defined feature, in one dimension there is a much smoother onset of the correlations. We will return to this point in Section~\ref{sec:T_dep}.

In order to use the decorrelator in a meaningful way at a given, common, temperature for both copies, I and II, of the system, we need to make sure that the energy difference of the two copies due to the local perturbation is consistent with arising from a typical thermal fluctuation. To ensure this, the energy difference should be small compared to the temperature:   
\begin{align}
    |\delta E|=\left|\sum_{j\in i=0}J_{ij}\delta{\bf S}_{i=0}\cdot{\bf S}_j\right|\sim J\varepsilon \ll k_B T\ .
\end{align}

This condition is fulfilled with the following order of limits
\begin{equation}
 \lim_{T\rightarrow 0}\lim_{\varepsilon\rightarrow 0}. 
 \label{eq:orderlimits}
\end{equation}
This is done in our calculations for both the linearised and non-linear equations \cite{PhysRevLett.121.024101}.


\subsubsection{Precessional dynamics and effective hydrodynamics}
We are interested in the real-time many-body dynamics of a spin system initialised in a state representative of a particular temperature. 
The dynamics of the classical spin system is generated by the spin-Poisson bracket
$$\{f,g\}=\sum_k\epsilon^{abc}\frac{\partial f}{\partial S^\alpha_k}\frac{\partial g}{\partial S^\beta_k}S^\gamma_k $$ 
where $f$ and $g$ are functions of the spins. This leads to the equation of motion 
\begin{align}
\partial_t{\bf S}_i=J \sum_j {\bf S}_i\times {\bf S}_j
\label{eq_eom}
\end{align}
which is just the precession of the spins in the local exchange field due to its neighbours.

The dynamics of the classical Heisenberg and related models have been extensively studied in the literature both numerically \cite{Spin_dyn_review,PhysRevLett.60.2785,PhysRevB.87.075133,PhysRevB.49.3266,2D_FM_dyn,2D_AFM_dyn} as well as using hydrodynamic approaches \cite{PhysRev.188.898,das2020nonlinear}. At high temperatures spin-spin dynamical correlators are generally found to be diffusive \cite{PhysRev.188.898,PhysRevLett.60.2785,PhysRevB.87.075133} whereas below the transition the dynamical spin structure factor reveals characteristic spin wave features both in 3D with a true thermodynamic phase transition \cite{PhysRevB.49.3266}, and as a crossover in 2D \cite{2D_FM_dyn,2D_AFM_dyn}. For 3D, in particular, the hydrodynamic theory of spin waves  \cite{PhysRev.188.898} predicts that for the Heisenberg ferromagnet at long wavelengths, the spin-wave dispersion is given by
\begin{align}
    \omega_{\bf k}=\frac{\rho_s}{m_\mathrm{ord}}|{\bf k}|^2-iw |{\bf k}|^4
    \label{eq_fmhyd}
\end{align}
where $\rho_s$ is the spin stiffness constant, $m_\mathrm{ord}$ is the equilibrium magnetisation and $w$ is the strength of spin-wave scattering. For the antiferromagnet, the low energy spin-waves in hydrodynamics obey
\begin{align}
    \omega_{\bf k}=\sqrt{\frac{\rho_s}{\chi_s}}|{\bf k}|-iD|{\bf k}|^2
    \label{eq_afmhyd}
\end{align}
where $\chi_s$ and $D$ denotes the spin susceptibility and diffusion constant respectively. The ballistic propagation of the linear spin-waves is correctly captured within the spin-wave theory as shown in the Appendix \ref{appen_sw}. At this point we note that at least in one dimension the long-time behaviour of the dynamic correlations may be dictated by non-linear effects and need to be studied within the theory of non-linear fluctuating hydrodynamics \cite{PhysRevLett.111.230601}.

{Above $T_c$, the conservation of the total magnetisation $\vect{S}_T$ (Eq.~\ref{eq_stotal}) in the symmetric phase and energy $E$ implies that for this short range interacting system both these currents are conserved leading to corresponding diffusion equations}, with diffusion constants $D_s$ and $D_E$ respectively. The diffusion of these conserved quantities then captures the long-time hydrodynamics of the high temperature paramagnetic phase \cite{kadanoff1963hydrodynamic,bloembergen1949interaction,PhysRevB.9.2171}.

In the following, we will in particular discuss how this `standard time-ordered' diffusive dynamics is related to the dynamics of an OTOC.


\subsubsection{Dynamics of difference field}
The decorrelator directly depends on the dynamical evolution of the difference field $\delta{\bf S}_{i}={\bf S}_{i}^{\mathrm{II}}-{\bf S}_{i}^{\mathrm{I}}$ defined above.  Using the equation of motion (Eq.~\ref{eq_eom}) and writing ${\bf S}_{i}^{\mathrm{II}}={\bf S}_{i}^{\mathrm{I}}+\delta{\bf S}_{i}$ (Eq.~\ref{eq_dels}), we obtain the equation of motion for the difference, $\delta{\bf S}_{i,t}$, as 
\begin{align}
\partial_t\delta {\bf S}_i=&
J{\bf S}_i\times\sum_{j\in i}\delta{\bf S}_j+
J\delta{\bf S}_i\times\sum_{j\in i}{\bf S}_j\nonumber\\
&+J\delta{\bf S}_i\times\sum_{j\in i}\delta{\bf S}_j.
\label{eq_deltas}
\end{align}
where we have dropped the superscript for clarity,  ${\bf S}^{\mathrm{I}}_{i}(\equiv {\bf S}_{i})$. This shows that the $\delta{\bf S}_i$ evolves in the background of the {dynamic spin field ${\bf S}_{i}$.}

In the numerical simulations we will mainly consider the limit $\varepsilon \rightarrow 0$, such that the second line in the equations of motion for the difference field drops out, and all quantities in the simulations are defined without factors of $\varepsilon$. This we refer to as the `linearised' decorrelator, which is not to be confused with the linear spin-wave analysis: the former preserves the chaotic nature of the dynamics, while the latter does not.

\section{Chaos at low temperature}
\label{sec:lowT}

We start our analysis of spatiotemporal chaos at low temperatures where spin rotation symmetry is spontaneously broken in an ordered state.
This goes along with the emergence of long-lived Goldstone mode. The items we discuss are existence and nature of a`short-time' integrable regime of non-interacting wavepacket propagation and a long-time `hydrodynamic' chaotic regime. A central finding is the fact that the Lyapunov exponent of this chaos is directly related to the spin-wave scattering rate. In between those two regimes, the scarred regime mentioned above appears.  

For a ferromagnet (antiferromagnet on a bipartite lattice), the ordering in question is uniform (N\'eel) order. The following discussion is centred on the ferromagnetic case. The analysis for the antiferromagnet, which is largely analogous albeit somewhat more complicated in terms of calculations, is relegated to the Appendix~\ref{appen_afmdecor}.

As remarked above, despite the lack of true long-range order in thermodynamic limit for $d=1,2$, for finite systems the rapidly
increasing correlation length at low temperature leads to the effective appearance of local order capable of supporting long-lived elementary excitations. Our discussion therefore proceeds in terms of spin waves about an {\it ordered} background, appropriate at least for time scales short compared to the spin-wave life- and scattering time and length-scales smaller than the correlation length.

\subsection{Equation of motion at low T}

Our approach is to cast the equations of motion, Eq.~\ref{eq_deltas} into the form of a linear term corresponding to the propagation of an `integrable' wavepacket, subject to non-linear `scattering' processes. The analysis of the latter will underpin our central result linking the spin-wave lifetime and the Lyapunov exponent, Eq.~\ref{eq_lambdatau}.

Deep inside the symmetry broken phase, we can expand in small deviations from the collinear ordering pattern,
\begin{align}
{\bf S}_i={\bf n}_i\sqrt{1-{\bf L}_i^2}+{\bf L}_i
\label{eq_spin_wave}
\end{align}
where ${\bf n}_i$ represents the direction of collinear order  whereas ${\bf L}_i$ is the spin-wave amplitude (with ${\bf n}_i\cdot{\bf L}_i=0$) and the latter are the gapless Nambu-Goldstone modes describing the lowest energy long wavelength excitations about the ground state. For the ferromagnet (antiferromagnet) they have quadratic, $\omega\sim k^2$ (linear, $\omega\sim k$) dispersion (Appendix \ref{appen_sw}). This suggests that the low energy dynamics of the difference field, $\delta{\bf S}_i$, is best understood in terms of the interaction of spin waves.  To this end we use Eq.~\ref{eq_spin_wave} in the evolution equation (Eq.~\ref{eq_deltas}), to obtain
\begin{align}
\partial_t\delta{\bf S}_i=&{\bf n}_i\times\sum_{j\in i}J_{ij}\delta{\bf S}_j+\delta{\bf S}_i\times\sum_{j\in i}J_{ij}{\bf n}_j\nonumber\\
&+{\bf L}_i\times\sum_{j\in i}J_{ij}\delta{\bf S}_j+\delta{\bf S}_i\times\sum_{j\in i}J_{ij}{\bf L}_j\nonumber\\
&-\frac{1}{2}{\bf L}_i^2{\bf n}_i\times\sum_{j\in i}J_{ij}\delta{\bf S}_j-\frac{1}{2}\delta{\bf S}_i\times\sum_{j\in i}J_{ij}{\bf n}_j~{\bf L}_j^2\nonumber\\
&+\delta{\bf S}_i\times\sum_{j\in i}J_{ij}\delta{\bf S}_j
\label{eq_ds_sw_2}
\end{align}
where we have used $\sqrt{1-{\bf L}_i^2}\approx1-\frac{1}{2}{\bf L}_i^2$, as appropriate for low temperatures.

For the nearest neighbour ferromagnet, $J<0$, we can choose without loss of generality 
 ${\bf n}_i=\hat{\bf z}$ (and hence ${\bf L}\perp \hat{\bf z}$). Therefore from Eq.~\ref{eq_ds_sw_2}, we get, after Fourier transforming (see Appendix \ref{appen_ferrodecor} for details), 
\begin{align}
    \partial_t\delta{\bf S}_{{\bf k}}=&\gamma_{\bf k}\mathcal{Z}\cdot\delta{\bf S}_{{\bf k}}+\frac{1}{N}\sum_{\bf q}\mathcal{A}_{\bf k,q}\cdot\delta{\bf S}_{\bf k-q}\ .
    \label{eq_eom_mct_fm}
\end{align}
For the first linear term in $d=1,2,3$ dimensions
\begin{align}
    \gamma_{\bf k}=2|J|\left(d-\sum_{i=1}^d\cos k_i\right)
    \label{eq_ferro_disp}\ .
\end{align}
and the matrix $\mathcal{Z}$  is given by Eq.~\ref{eq_zmat}. This
term corresponds to the linearised equations of motion and accounts for the free ballistic propagation of $\delta{\bf S}_{\bf k}$.

The non-linear second term represents the scattering of $\delta{\bf S}_{\bf k}$ with the dynamic spin-waves with the scattering determined by the matrix
\begin{align}
    \mathcal{A}_{\bf k,q}=\mathcal{O}_{\bf k,q}+\mathcal{M}_{\bf k,q}\ .
    \label{eq_fmvertex}
\end{align}
While the detailed forms are given by Eq.~\ref{eq_fmo} and \ref{eq_fmmkq}, we note that these scatterings imply that the different Fourier modes of difference field scatter from the dynamic spin-wave modes and this results in the mode-coupling route to chaos at low temperatures as we shall see below. From the explicit forms of the scattering kernels, it is clear that the scattering vanishes as ${\bf k}\rightarrow 0$ and hence the long-wavelength modes are more long-lived, as expected for Goldstone modes. 

\subsection{The linearised de-correlator and emergent integrability}
\label{ssec:freesolution}

 The full solution can be expressed in integral form as 
\begin{align}
\delta {\bf S}_{\bf k}(t)=\delta {\bf S}_{\bf k}^{0}(t)+G^{0}_{\bf k}(t)\sum_{\bf q}\int_0^t dt'~\mathcal{A}_{\bf k,q}(t')\cdot\delta {\bf S}_{\bf k-q}(t')
\label{eq_mctfm}
\end{align}
This can then clearly be used as the starting point of {the mode-coupling \cite{reichman2005mode} expansions for the difference field at low temperatures.} 

The free solution is obtained by setting $\mathcal{A}_{\bf k,q}=0$. This is given by the first term of Eq.~\ref{eq_mctfm} as
\begin{align}
    \delta{\bf S}^{0}_{{\bf k}}(t)=G^{0}_{{\bf k}}(t)\cdot\delta{\bf S}_{{\bf k}}(0)
    \label{eq_luv}
\end{align}
where the free propagator $G^{0}_{{\bf k}}(t)$ is given by Eq.~\ref{eq_free_propagator}. In this case each momentum mode is independent. However they interact with different spin-wave modes eventually leading to the coupling of different modes of decorrelations. We now first develop the {form of the decorrelator from the} free solution followed by inclusion of  scattering further down.

The explicit form of the free solution is obtained from Eq.~\ref{eq_luv} and is thus
\begin{align}
\delta{\bf S}_{\bf k}=\epsilon^z\hat{\bf z}+\eta\left[\cos\left(|\gamma({\bf k})|t+\phi\right)\hat{\bf x}+\sin\left(|\gamma({\bf k})|t+\phi\right)\hat{\bf y}\right]
\label{eq_deltas_SW}
\end{align}
where $\eta=\sqrt{(\epsilon^x)^2+(\epsilon^y)^2}=\varepsilon \sqrt{1-|{\bf L}_0(0)|^2}$ and hence proportional to magnetisation and 
$\phi=\tan^{-1}(L^y(0)/L^x_0(0))$. Thus in the thermodynamic limit, we have (see Appendix \ref{appen_ferrodecor} for details)
\begin{align}
    \frac{\mathcal{D}(i,t)}{\varepsilon^2}=(1-m_T^2)^2\delta_{i,0}+m_T^2\mathcal{F}_d(i,t)
    \label{eq_zerodit_22_main}
\end{align}
where $m_T$ is the magnetisation and  $\mathcal{F}(i,t)$ is a dimension dependent function given by
\begin{align}
    \mathcal{F}_{d}(i,t)=\left\{\begin{array}{l}
\left(J_{i_1}(2t)\right)^2~~~~~~~~~~~~~~~~~~~~~~~~~~~~~~~~~~~~~~~~~(d=1)\\
\left(J_{i_1}(2t)\right)^2\left(J_{i_2}(2t)\right)^2~~~~~~~~~~~~~~~~~~~~~~~~(d=2)\\
\left(J_{i_1}(2t)\right)^2\left(J_{i_2}(2t)\right)^2\left(J_{i_3}(2t)\right)^2~~~~~~~(d=3)\\
\end{array}\right.
\label{eq:FM_LSW_sol}
\end{align}
with $i\equiv(i_1,i_2,\cdots i_d)$ denotes the position on a $d=1,2,3$ dimensional cubic lattice and $J_\nu(t)$ denotes Bessel function of first kind.

\begin{figure}
    \centering
    \includegraphics[width=0.475\textwidth]{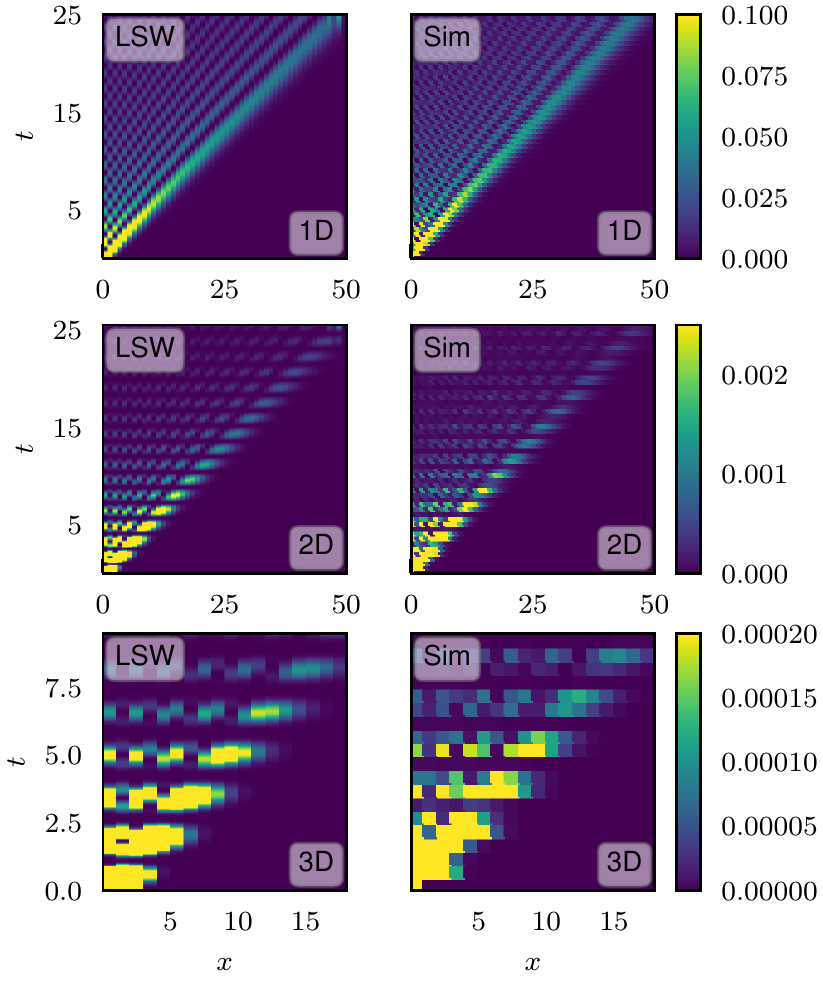}
    \caption{Spatio-temporal behaviour of  the linearised de-correlator for the nearest neighbour ferromagnet in $d=1,2,3$ (top to bottom) given by Eq.~\ref{eq_zerodit_22_main} computed from linear spin wave (LSW) (left panels), compared to the full numerical soution (right panels). The de-correlator spreads ballistically, but does not grow exponentially in the non-interacting solutions and not appreciably in the interacting solutions at these short times.}
    \label{fig_fmzero}
\end{figure}

We plot $\left[\mathcal{D}(i,t)/\varepsilon^2-(1-m_T^2)\delta_{i0}\right]/{m_T^2}$ from Eq.~\ref{eq_zerodit_22_main} for a one-dimensional cut along the cartesian coordinate axes in dimensions $d=1,2,3$ in the left panels of Fig. \ref{fig_fmzero}, and compare to the corresponding results of the numerical simulations (see \ref{app:numerics}) of the full spin dynamics in the right panels. The striking similarity of the two provides crucial evidence of the emergent integrability at low temperatures in the symmetry broken phase. We note that the time and length scales shown here are chosen to be below the inverse Lyapunov time, $\lambda^{-1}$ and below the correlation length, $\xi^{-1}$, and, thus the approximation of non-interacting spin waves in an ordered background is expected to be good as potential scattering between spin waves will be negligible.

At this linear order the solution in Eq.~\ref{eq_deltas_SW} has the same structure as the linear spin-wave solution (Appendix \ref{appen_sw}), but with a localised (in real space) initial condition. Therefore the short-time behaviour of the de-correlator is nothing but akin to the spread of an initially localised spin-wave packet. This, in the linear-spinwave regime is a ballistic phenomena without any exponential amplification in conformity with the full numerical calculation. We further note the dimensionality dependence of the striking beating structures visible in Fig. \ref{fig_fmzero}, and also reflected in the full numerical dynamics, expected for interference of non-interacting spin waves. 

\begin{figure}
    \centering
    \includegraphics[scale=0.9]{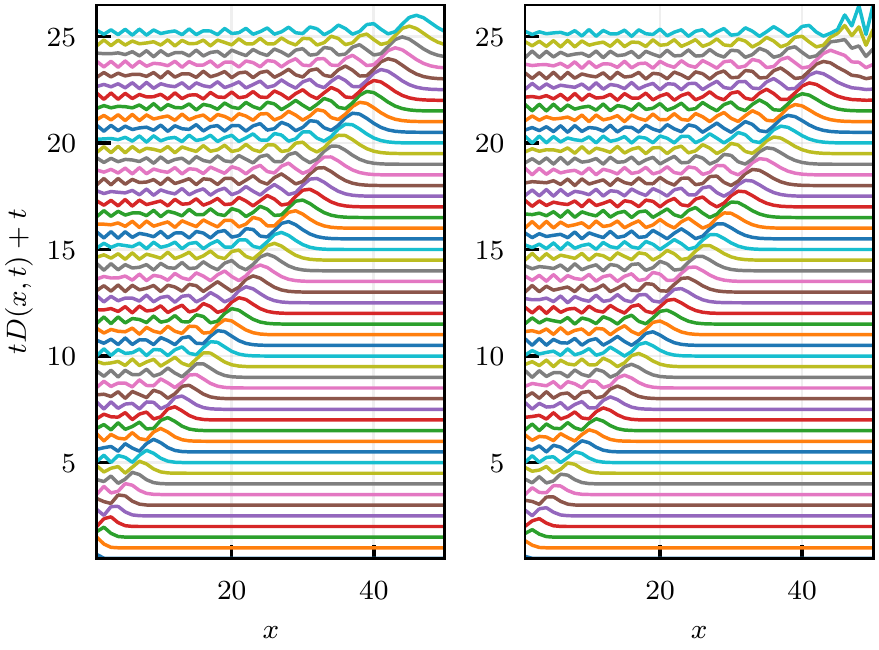}
    \caption{Constant time-slice plots for the linearised decorrelator in the full simulations (left) and as  given by Eq.~\ref{eq_zerodit_22_main} (right) for $d=1$ nearest neighbour ferromagnet. Slices are scaled by t and offset vertically by t for visual clarity, and plotted for $t=0.5,\cdots,25$ in steps of 0.5.
    \label{fig_tslicefm}
    }
\end{figure}

We provide a more detailed comparison of the above short-time physics in the one-dimensional case  via constant time slices plotted in Fig.~\ref{fig_tslicefm}. Besides the fine-structure visible within the emergent light-cone of the decorrelator, we also observe a propagating peak -- {\it the primary packet of decorrelation}-- travelling out at twice unit speed (in units of $|J|$ which has been chosen to be unity in Fig. \ref{fig_fmzero} and the rest of this work). Notice that this is also the case at $T=\infty$, however the mechanism to generate the chaos and the butterfly velocity is different as there are no spin waves at infinite temperature as shown by the analysis of the two-point spin correlators which are purely diffusive.

Finally by using the asymptotic form of the Bessel function given by
\begin{align}
    J_\nu(t)\sim\sqrt{\frac{2}{\pi t}}\cos\left(t-\frac{\nu\pi}{2}-\frac{\pi}{4}\right)
    \label{eq_asymfm}
\end{align}
we obtain powerlaw decay $\sim t^{-d}$ of Eq.~\ref{eq_zerodit_22_main} in agreement with the full numerical simulations as an intermediate asymtotic behaviour until scattering become important. 
To understand the ballistic nature and the emergence of a light-cone, one needs to consider the asymptotic scaling along rays of fixed velocity $v=x/t$. Using stationary phase arguments, see App.~\ref{app:LSW_asymptotics}, this can be seen to result in power-law decay for $v_i \le  2J$, and exponential decay for $v_i>2 J$. This sharp distinction then defines the butterfly speed $v_b$, and is the origin of the light-cone observed above.

This free solution is expected to properly capture the short time behaviour ($t\ll\lambda^{-1}$), where no significant exponential growth of the decorrelator has taken place and hence $\lambda\approx0$.
In Sec.~\ref{sec:T_dep} and Fig.~\ref{fig:lambda_vs_T}, we demonstrate that the time-window for the validity of the linear solution expands and tends to diverge as the temperature is decreased to zero. 
We provide a more detailed picture of this short time integrability at short times and low temperatures via constant space slices in app.~\ref{appen_ferrodecor}, Fig. \ref{fig_xslice_appen}.

\subsection{Long-time chaos through scattering with spin-waves}
\label{ssec:longtime}

We now turn to the effects of the spin waves, ${\bf L}$, scattering with the difference field, $\delta{\bf S}_i$. 
At times much longer than the Lyapunov time, $\lambda^{-1}$, chaos sets in which is expected to arise due to the scattering of spin waves. Indeed, we establish that it is the scattering time of the spin waves which directly determines the Lyapunov exponent. 

At the outset we note that the light-cone velocity remains unchanged even at late times when scattering cannot be neglected. This can be understood as follows. As the packet of $\delta{\bf S}_i$ propagates outward it will eventually scatter off spin-waves in the background field, either being reflected back inside the light-cone or splitting into two spin-waves propagating forwards along the light-cone  and backwards inside the light-cone respectively. Thus, the leading edge, determining the light-cone velocity, will always be dominated by the remaining weight of the initial peak that was never reflected propagating outwards at the initial velocity given by Eq.~\ref{eq_zerodit_22_main}. In other words infinitesimally near the light-cone edge, the linear form of the de-correlator in Eq.~\ref{eq_zerodit_22_main} always remain valid and hence correctly predicts the light-cone velocity.

\begin{figure}
    \centering
    \includegraphics[scale=0.95]{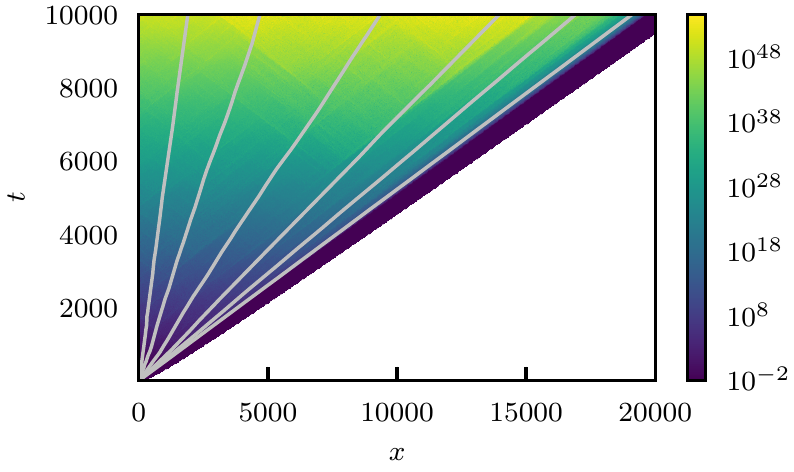}
    \caption{Late time light-cones. Spatio-temporal behaviour of the linearised decorrelator $D(x,t)$ with logarithmic colour-scale. Light gray lines are logarithmic contours, defined from $i_m(t)$ with $\sum_{i=0}^{i_m(t)}$$\log(D(i,t)/ \sum_i \log(D(i,t))$ $> k$ for  $k=$ 0.1, 0.25, 0.5, 0.75, 0.9, 0.99. 1D ferromagnet on a system of size $L=40000$ at $T=0.04$.
    \label{fig:FM_LC_long_times}}
\end{figure}

To demonstrate this, we show the late time light-cone for the one-dimensional ferromagnet in Fig.~\ref{fig:FM_LC_long_times} for a system with $L=40000$ at $T=0.04$ for times up to $t=20000$. We note that here the inverse Lyapunov time is $t_{\lambda}=1/\lambda \approx 170$, thus, we are probing the dynamics deep into the chaotic temporal regime (see below), and for spatial scales considerably larger than the spin-spin correlation length $\xi \approx 20$.
Remarkably, we still observe a perfectly linear light-cone with a velocity determined by the spin wave velocity as for the short time dynamics discussed above.

\begin{figure}
    \centering
    \includegraphics[]{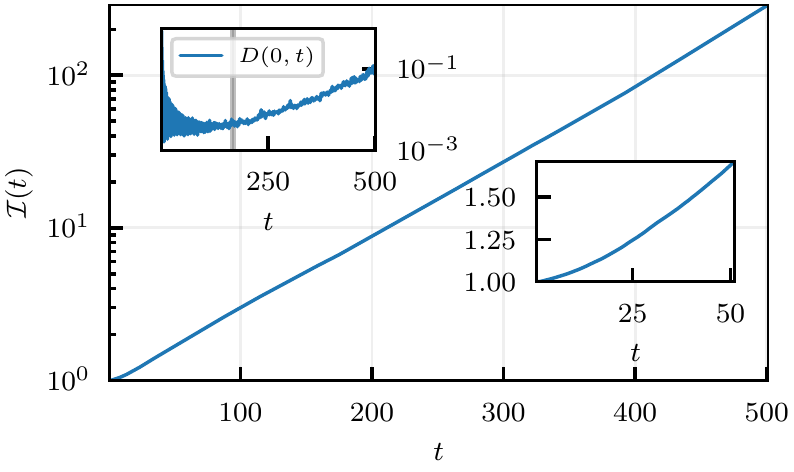}
    \caption{Onset of chaos. Temporal behaviour of the linearised decorrelator for the FM in $d=1$ shown for $T=0.04$ and $L=1000$. Main panel: Summed decorrelator $\mathcal{I}(t)$ versus time $t$ on a log-linear scale demonstrating exponential growth $\mathcal{I}(t) \sim e^{2 \lambda t}$. Upper left inset: Decorrelator at the inital site $D(i=0,t)$ on a log-linear scale showing initial power law decay $D(i=0,t) \sim 1/t$ up to a time $t_{\lambda} \sim 1/\lambda\approx 170$ (indicated by gray vertical line) followed by the same exponential growth $D(i=0,t) \sim e^{2 \lambda t}$. Lower right inset: Initial time behaviour of the summed decorrelator on a linear scale showing slow initial growth for time $t \ll 1/\lambda$.
    \label{fig:FM_D_cor}
    }
\end{figure}

However well within the light-cone repeated scattering results in the rapid amplification of the signal at times greater than $\lambda^{-1}$ due to the non-linear contributions denoted by the mode-coupling term in Eq.~\ref{eq_eom_mct_fm}. This temporal growth is most directly seen in the space averaged de-correlator, {\it i.e.}, $\mathcal{I}(t)$ as given by Eq.~\ref{eq_sumdecorexp}. In fact, it is possible to show from  \ref{eq_zerodit_22_main} that within linear theory $\mathcal{I}(t)=\varepsilon^2$, {\it i.e.}, a constant. Thus, any deviation, in particular, an exponential growth of this, is a direct indicator of the chaotic  regime where scattering is important. This is shown in the main panel of Fig. \ref{fig:FM_D_cor} which for long times clearly exhibits exponential growth, $\mathcal{I}(t)\sim e^{2\lambda t}$ with the Lyapunov exponent, $\lambda$. Indeed, we generally find that the Lyapunov exponent based on the summed decorrelator converges to the exponential form on times of order $1/\lambda$.  

In contrast, as mentioned above, the upper left inset shows the decorrelator at the initial site of perturbation, {\it i.e.}, $D(i=0,t)$ where the initial $t^{-1}$ powelaw decay is clearly visible till $t_{\lambda} \sim 1/\lambda$. After $t^*$ the exponential growth takes over with the same Lyapunov exponent, $\lambda$, with which $\mathcal{I}(t)$ grows. 

Finally, the lower right inset shows the short time behaviour of the summed decorrelator on a linear scale, showing only slow initial growth, which a-posteriori justifies our treatment of spin-waves as (almost) non-interacting in this regime. 

\subsubsection{Mode-coupling theory and the emergence of chaos}

The mode-coupling theory for $\delta{\bf S_k}$ introduced above (Eq.~\ref{eq_mctfm}) can provide crucial insight into the temporal aspects of the {\it late-time chaos} in terms of the properties of the spin-waves. This section is devoted to the derivation of the result linking the Lyapunov exponent, 
to the spin-wave scattering rates via Eqs.~\ref{eq_lambdatau} and \ref{eq:lyapunov}.
 
Our analysis starts by expanding the integral Eq.~\ref{eq_mctfm} iteratively to obtain (using Eq.~\ref{eq_luv} and Eq.~\ref{eq_sk_flat}) a mode coupling expansion of $\delta{\bf S}_{\bf k}(t)$. The details are given in Appendix \ref{appen_ferrodecor}. From this, we readily obtain the leading order contribution to the space-averaged de-correlator (Eq.~\ref{eq_sumdecorexp}) given by
\begin{widetext}
\begin{align}
\delta {\bf S}_{\bf k}(t)\cdot\delta{\bf S}_{\bf -k}(t)=&\epsilon^2+\frac{1}{N}\boldsymbol{\epsilon}^T\cdot\sum_{\bf q_1}\int_0^tdt_1\left[\mathcal{A}_{\bf -k;q_1}(t_1)G^0_{\bf -k-q_1}(t_1)+ \left[G^0_{\bf k-q_1}(t_1)\right]^T\left[\mathcal{A}_{\bf k;q_1}(t_1)\right]^T\right]\cdot\boldsymbol{\epsilon}+\cdots
\label{eq_mctsummedfm}
\end{align}
\end{widetext}
where $\cdots$ refer to higher order terms. The first term is indeed the constant contribution of the free decorrelator as discussed above.

The first order correction to this is due the scattering as denoted by the second term on the right hand side of the above expression. This scattering term contains two separate contributions corresponding to the two terms in Eq.~\ref{eq_fmvertex}. For the first contribution which is proportional to
\begin{align}
    \left[\mathcal{O}_{\bf -k;q_1}(t_1)G^0_{\bf -k-q_1}(t_1)+ \left[G^0_{\bf k-q_1}(t_1)\right]^T\left[\mathcal{O}_{\bf k;q_1}(t_1)\right]^T\right],
\end{align}
explicit calculation shows that this term is proportional to $\epsilon^\alpha\epsilon^\beta$ for $\alpha\neq\beta$. This stems from the general antisymmetric structure of $\mathcal{L}_{\bf q}$ (Eq.~\ref{eq_lfm}). Hence on taking the average over the thermalised initial condition this term vanishes, since $\langle\epsilon^\alpha\epsilon^\beta\rangle_T\propto\delta_{\alpha\beta}$.

However, the second term gives a non-zero contribution of the form 
\begin{widetext}

\begin{align}
    \frac{\varepsilon^2m_T^2}{2N^2}\sum_{{\bf q}_1,{\bf q}_2}\int_0^t~dt_1~\left({\bf L}_{{\bf q}_1}\cdot{\bf L}_{{\bf q}_1-{\bf q}_2}\right)\left[(\gamma_{{\bf k}-{\bf q}_1}-\gamma_{{\bf q}_1})\sin(|\gamma_{{\bf k}-{\bf q}_1}|t)+(\gamma_{{\bf k}+{\bf q}_1}-\gamma_{{\bf q}_1})\sin(|\gamma_{{\bf k}+{\bf q}_1}|t)\right]
    \label{eq_mfirst}
\end{align}
\end{widetext}
 which then is of the same form as the second term in Eq.~\ref{eq_zerodit_22_main} albeit summed over the lattice points. We note that in deriving the above expression we have not assumed free-spin waves. 

At second order, there are three classes of terms which can be schematically written as $\mathcal{O}\mathcal{O}$,  $\mathcal{O}\mathcal{M}$ and $\mathcal{M}\mathcal{M}$. Again due to the averaging the cross terms vanish. While the third term is nothing but the higher order version of Eq.~\ref{eq_mfirst}, the first term is the sub-leading contribution with respect to Eq.~\ref{eq_mfirst}. This provides the basis for neglecting all terms proportional to $\mathcal{O}$. The dynamics of the difference field (Eq.~\ref{eq_eom_mct_fm}) now becomes exactly equivalent to that of interacting spin-waves for a ferromagnet in Eq.~\ref{eq_spinwave_nonlinear}. 

{This concretely shows that the same coupling of the modes that gives rise to the spin-wave lifetime also leads to chaos. The respective forms are obtained by re-summing the series in Eq.~\ref{eq_mctsummedfm} within the above approximation (and similar series for the spin-waves). However, note that, in the regime where spin wave scattering leads to chaotic behaviour, the individual modes of the exponentially growing difference field, $\delta{\bf S}_i$, cease to be well-defined in the long-time limit. It is therefore not possible to follow the exponential growth of a particular ${\bf k}$ mode in isolation from that of the others. Then, it is the combined effect of all interacting modes which is measured in the summed decorrelator, and its form is suggestively written as 
\begin{align}
    \mathcal{I}(t)\sim\sum_{\bf k} e^{2t/\tau_{\bf k}}\ .
    \label{eq_lambdatau}
\end{align}
where $\tau_{\bf k}$ is the lifetime of the ${\bf k}$th mode.}

{At long times the actual Lyapunov exponent defined by $\mathcal{I}(t)\sim e^{2\lambda t}$ will therefore be set by the lifetime of the short-lived spin waves:
\begin{align}
   \lambda\sim\max_{\bf k}\frac{1}{\tau_{\bf k}} 
   \label{eq:lyapunov}
\end{align}
}
Recall that the limit of $\epsilon\rightarrow 0$ has been taken in the numerical calculations to open a sufficiently long time-window (in fact infinitely long for the linearised calculations as described in App.~\ref{app:numerics}) before the decorrelator saturates due to the finite phase-space volume of the unit sphere.

The above mode coupling theory therefore connects the chaos time-scale and the lifetime of the quasiparticles in a classical system. Similar results have been proposed recently in context of quantum many-body systems, particularly Fermi-liquids \cite{AltmanScrambling2020}.

We emphasize that $\lambda$ is then dominated by the short scattering time-scales which are necessarily away from the ordering momentum, ${\bf k}=0$, where $\tau_{\bf k}$ diverges for the Goldstone modes and hence the present finite $\lambda$ is not inconsistent with the very long-lived Goldstone modes at the longest wavelengths.


\begin{figure}
    \centering
    \includegraphics[]{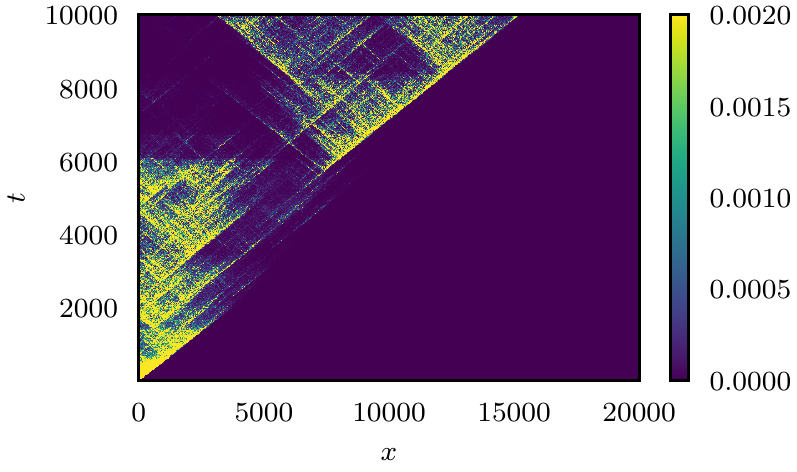}
    \includegraphics[]{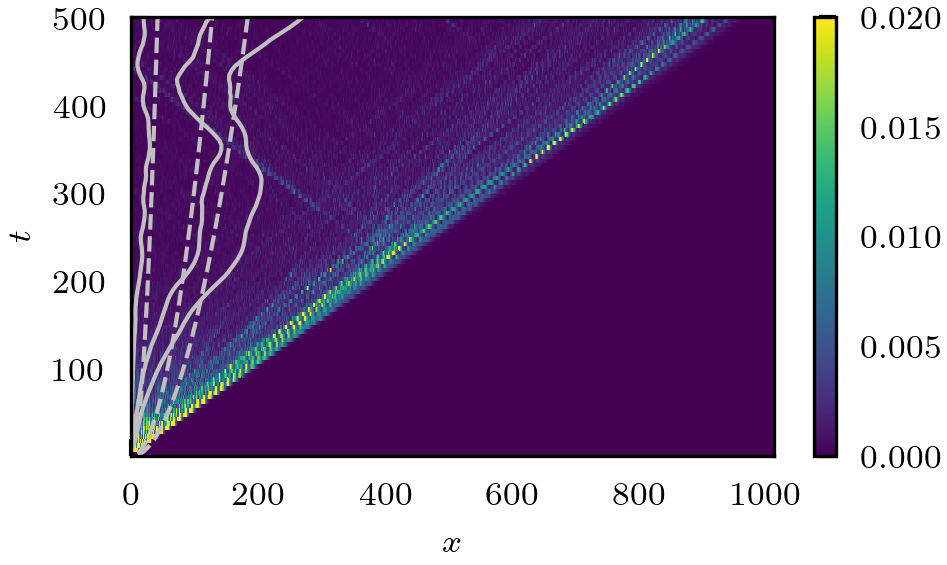}
    \caption{Scarred decorrelator of the 1D FM. (a) Spatio-temporal behaviour of the linearised normalised decorrelator $D(x,t)/\mathcal{I}(t)$ on a linear colourscale. Same parameters as in Fig.~\ref{fig:FM_LC_long_times}, $L=40000$ and $T=0.04$. (b) Same on smaller spatio-temporal scales including linear contours (solid lines) and fits (dashed lines) $x^2 \sim t$ to the contours assuming diffusive scaling.
    \label{fig:FM_LC_scarred}
    }
\end{figure}

\subsection{Cross-over between integrability and chaos: the scarred regime}
\label{ssec:crossscar}
We next address the nature of the cross-over between short-time integrable and the long-time chaotic regimes. 
Note that already in Fig.~\ref{fig:FM_LC_long_times}, some fine-structure is visible in the decorrelator at long times. To make this more visible we study the scaled decorrelator via $D_{scal}(i,t) = \mathcal{D}(i,t)/\mathcal{I}(t)$ which removes the exponential temporal growth and thus enhances the spatial structure at given time.
In Fig.~\ref{fig:FM_LC_scarred}, the scaled decorrelator exhibits a scarred appearance, due to the coexistence of a multitude of secondary light-cones at different spatial and temporal scales. We interpret these as originating at points where a given spin-wave with a well defined velocity and appreciable weight scatters of other spin-waves and get reflected, or "splits", resulting in a backward-propagating, or two spin waves propagating in opposing directions. Note that the spin wave can `survive' a scattering event, in the sense that it splits into a part which continues to propagate in its original direction, while a second part of the signal is seeded which propagates in the opposite direction. 

In turn, these products of the scattering event can individually scatter further, thus seeding further light cones; eventually, these overlap and merge, giving rise to the diffusive regime at the longest times. 
This has a natural interpretation in terms of the scars: with the increase in the number of lightcones, a given point in space receives contributions propagating outwards or inwards, whose statistics are those  of a random walk, whence the diffusive nature. 

\begin{figure}
\centering
\includegraphics[]{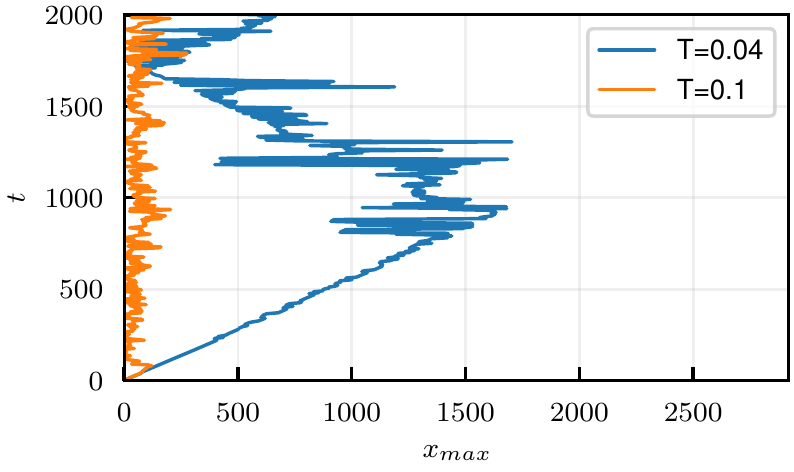}
\caption{The position of the maximum $x_{max}$ of the decorrelator versus time $t$ for $T=0.04$ and $T=0.10$. For short times spin-waves propagate freely, and the maximum tracks the light-cone velocity, whereas at later times spin-waves scatter. \label{fig_maxposscaled}}
\end{figure}

An alternate way to visualise the crossover is to track the maximum weight of the difference field in space-time plane; for the one dimensional ferromagnet, this is shown in Fig. \ref{fig_maxposscaled}. For short times, the maximum of the decorrelator tracks the lightcone. However at later times, the scattering events nucleate growth of the decorrelator and hence increase the number of subsequent scattering events, so that the maximum then moves closer to the origin, where more lightcones overlap. In Fig.~\ref{fig_maxposscaled}, this is clearly visible for the higher temperature $T=0.1$ where scattering is stronger. In contrast at $T=0.04$ we observe free propagation for a significantly longer time, and subsequently repeated scattering events during which the propagation direction of the main peak in the signal is fully reversed. This in turn supports that the life-time of the spin-waves is indeed larger than the scattering time, such that spin-waves undergo a random walk at fixed speed interrupted by reversal of the propagation direction when a scattering event occurs. Thus, in this regime we expect the emergence of a diffusive regime. 

We demonstrate this directly in the lower panel of Fig.~\ref{fig:FM_LC_scarred}, where we show contours to the scaled decorrelator and diffusive fits to the contour lines of the form $x^2 \sim t$. The results show large fluctuations, strongly enhanced by the chaotic growth, even for very large times due to the increasing life time of the spin waves at decreasing temperatures.

We note in passing that this cross-over is not the only manifestation of the existence of quasiparticles. There exists in addition a feature in the decorrelator at the initial site of the perturbation, which  shows powerlaw decay in  time for $t \lessapprox 1/\lambda$, before the exponential growth takes over. This is absent in the case of the kagome spin liquid discussed further down \cite{PhysRevLett.121.250602}.

At this point, we note that in practice we cannot probe the scarred behaviour in dimensions larger than 1 due to the very long time and system sizes required to make this scattering and the emergent diffusive core visible. 
\subsection{Bipartite Antiferromagnet}

\begin{figure}
    \centering
    \includegraphics[width=0.48\textwidth]{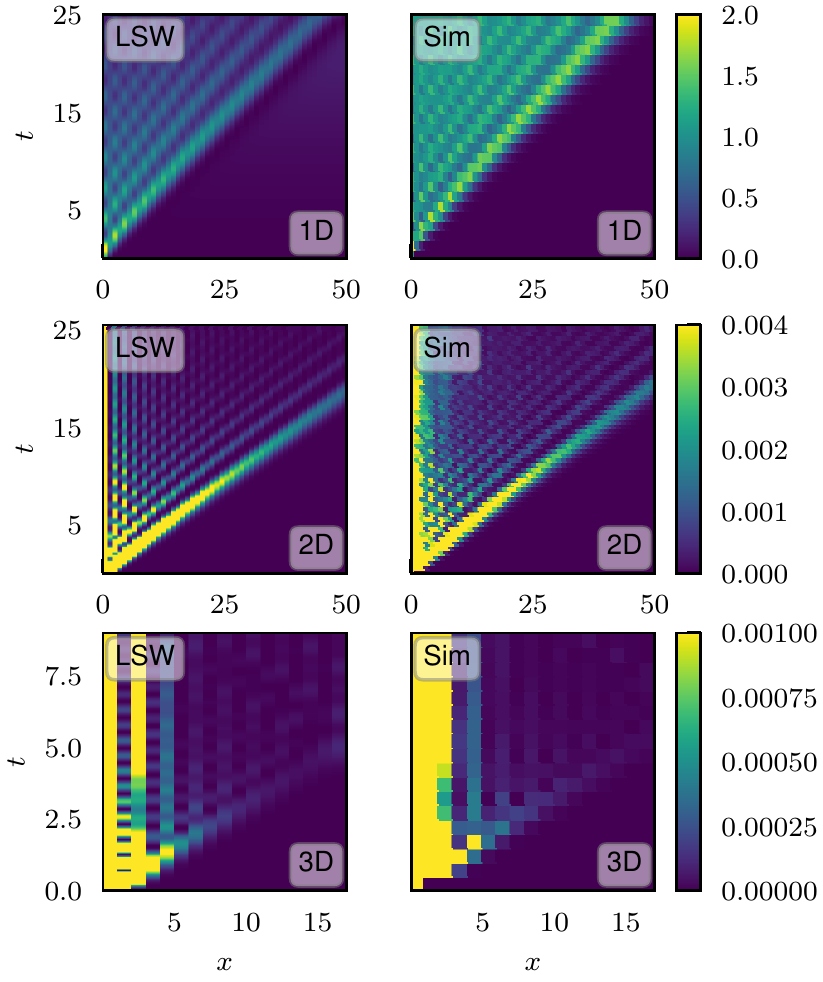}
    \caption{Spatio-temporal behaviour of  the linearised de-correlator for the AFM in $d=1,2,3$ (top to bottom) comparing the non-interacting spin wave theory (LSW) (left) to the full numerical dynamics (right) in the short time regime where exponential growth and spin wave scattering is negligible in the interacting dynamics.
    \label{fig:lcafm}
    }
\end{figure}
The situation for the nearest-neighbour bipartite antiferromagnet turns out to be quite close to that of the ferromagnet.  Nonetheless, there are some striking differences between the two, and we summarise the general features emphasising these differences in the following paragraphs. The details of the underlying calculations are relegated to Appendix \ref{appen_afmdecor} for the case of a Neel state,
\begin{align}
    {\bf n}_i=\left\{\begin{array}{l}
    \hat{\bf z}~~~~~~~~~~~~\forall i\in A\\
    -\hat{\bf z}~~~~~~~~~~\forall i\in B
\end{array}\right.
\label{eq_afmorder}
\end{align}
where $A$ and $B$ represent the two sublattices. 

The free decorrelator (expressions in Eqs. \ref{eq_afmdecor_freea} and \ref{eq_afmdecor_freeb}) is compared to the numerical solutions of the full spin dynamics in Fig. \ref{fig:lcafm} for short times. As in the ferromagnetic case, the two agree quantatively, both in the detailed spatiotemporal  interference patterns and the ballistically propagating peaks. We would like to note that for the antiferromagnet the free decorrelator at the origin-- the initial localtion of the difference field-- decays much more slowly compared to the ferromagnet, a behaviour which is expected from the long time scaling in the stationary phase calculation, see App.~\ref{app:LSW_asymptotics}.

%
However, the slow decay is soon swamped by the exponential amplification of the decorrelator as a whole. Indeed, these non-linear effects yield late time chaos like in the ferromagnet, albeit with different temperature dependence as we show below.

\subsection{The integrable regime, and cross-over to chaos, in \texorpdfstring{$d>1$}{}}

\begin{figure}
    \centering
    \includegraphics[scale=0.9]{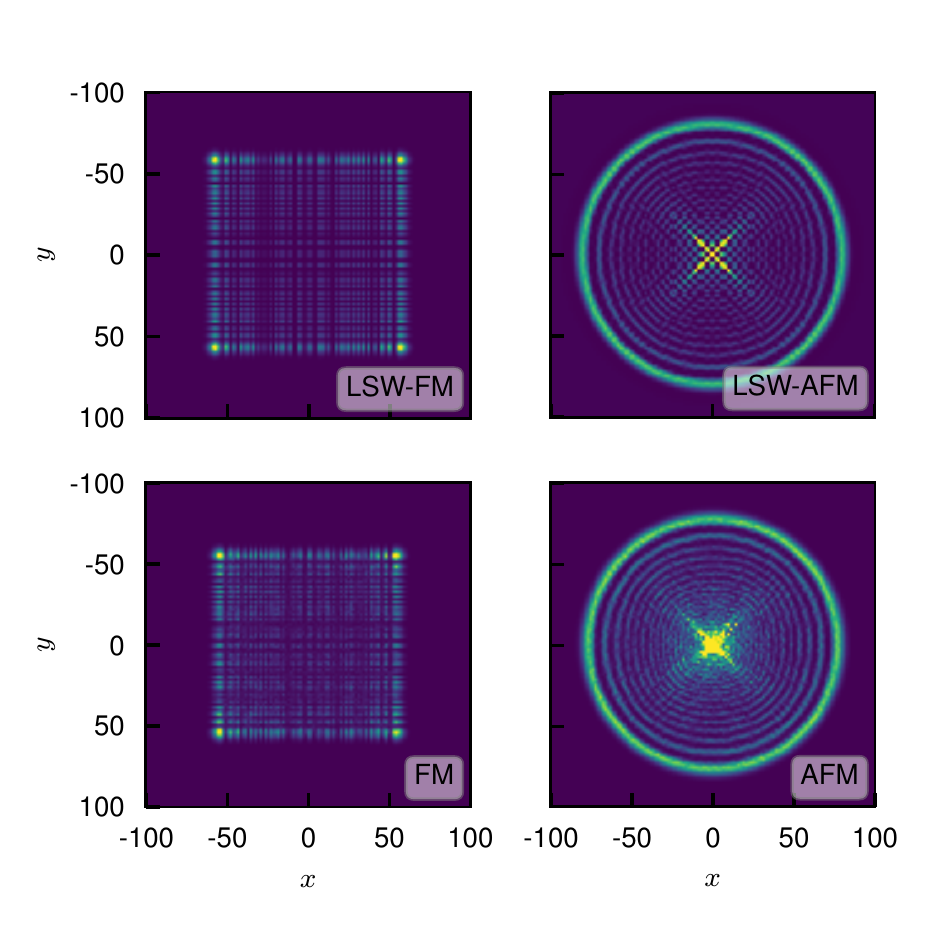}
    \caption{Shape of the wavefronts in 2D comparing the FM (left) to the AFM (right) and the linear spin wave results (top) to the full numerical simulations (bottom). All at t=30, numerical simulations for L=200 and T=0.1.}
    \label{fig:2D_wavefronts}
\end{figure}

The visually most striking difference between FM and AFM is the shape of the free decorrelator, plotted in  Fig.~\ref{fig:2D_wavefronts} for $d=2$ for LSW and for the full numerical solutions.
The FM (left panels) displays a square shape with bright peaks at the corner. This structure is readily inferred from the factorised form of Eq.~\ref{eq:FM_LSW_sol}. By contrast, the AFM (right panels) free decorrelator comes in the shape of an isotropic circle.

This difference is explained in terms of the origin of contributions to the difference field in $k$-space. For the ferromagnet, the explicit expression of the Bessel function as well as the stationary phase solution of the integral expression (Eq.~\ref{eq_zerodit_22_main_2}) shows that the integral is dominated by the weight of lattice dependent modes away from the ordering vector. This is related to the fact that the ferromagnetic spin waves have a quadratic dispersion. However for the antiferromagnet, the stationary phase solution (Eq.~\ref{eq_afmdecor_freea} and \ref{eq_afmdecor_freeb})  is dominated, due to the linearly dispersing spin waves, by the contribution from modes near the magnetic ordering vector. While we do not have straightforward closed asymptotic expression for the integral form, the isotropy nonetheless follows from the 
emergent rotational symmetry near these soft modes. Indeed a stationary phase approximation shows that the isotropic butterfly speed, $v_B=2J\sqrt{d}$, which matches very well with the numerical calculation as well as the full integral solution of the free decorrelator, App.~\ref{app:LSW_asymptotics}.

{This naturally raises the question about the persistence of  the square wavefront at long times for the nearest neighbour ferromagnet. We note that upon increasing temperature, the square gets replaced by an isotropic circular shape as the order underpinning the existence of the quasiparticles ceases.} 
We note in passing that for the classical spin liquid at low temperatures~\cite{PhysRevLett.121.250602}, the decorrelator is also circular. 

{Indeed, we cannot exclude that that nonlinear effects at long times could restore isotropy. If this were the case the chaotic long-time decorrelator at low temperatures would also become isotropic. A direct numerical verification of this is beyond our present numerical capacities even for $d=2$ on account of the rapidly growing length/timescales. At any rate, note that the decorrelator at these scales will be dominated by the exponentially growing core. This however only grows  diffusively, in contrast to the  ballistic spreading of the free decorrelator. A parametrically separated coexistence between a square form of the latter, and a circular one of the former, is hence another possibility.}

\section{Temperature dependence of chaos scales in \texorpdfstring{$d=1,2,3$}{} \label{sec:T_dep}}
We now collate our numerical results for the temperature dependence of the central chaos scales, the Lyapunov exponent and the butterfly velocity, for the hypercubic lattices in $d=1,2,3$, considering both the ferro- and the antiferromagnet.

\subsection{Lyapunov exponent}

\begin{figure}
    \includegraphics[scale=0.95]{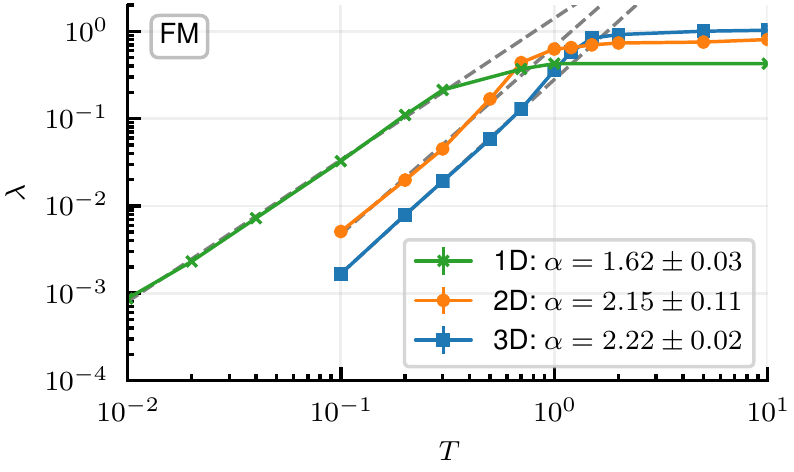}
    \includegraphics[scale=0.95]{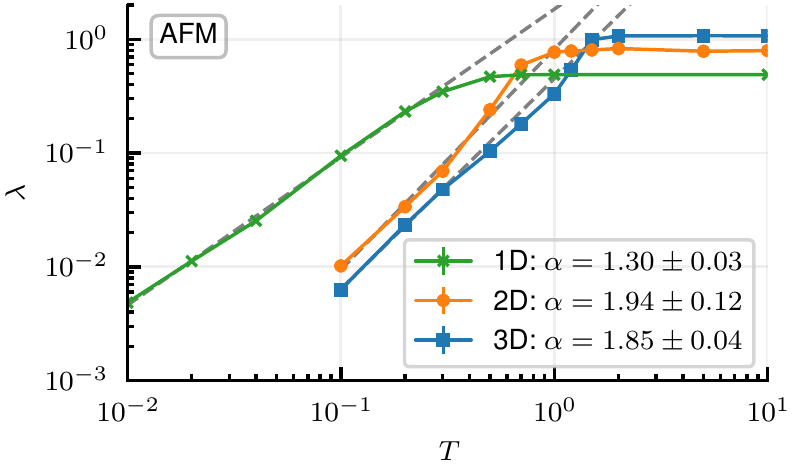}
    \caption{Temperature dependence of Lyapunov exponent for the FM (top) and the AFM (bottom). Fitting errors on the Lyapunov exponents are smaller than the symbol size. Dashed gray lines are power law fits to the low temperature behaviour, $\lambda(T) \sim T^{\alpha}$, with the best fit value and standard deviation given in the legend. Results are obtained for systems with $L=10000$, $L=400$, and $L=40$ in $d=1,2,3$ respectively. 
    \label{fig:lambda_vs_T}
    }
\end{figure}
The Lyapunov exponent,
Fig.~\ref{fig:lambda_vs_T}, was obtained from a fit to the time-dependence of the decorrelator. The data shown results from considering its behaviour at the initial site $D(x=0,t) \sim e^{\lambda t}$. We account for the initial powerlaw decay in the ordered regime (see Fig.~\ref{fig:FM_D_cor}) by only fitting after the cross-over time $t_{\lambda} \sim 1/\lambda$ where the behaviour is clearly exponential. We have corroborated these values by comparison to the spatially integrated decorrelator $\mathcal{I}(t) \sim e^{\lambda t}$, which does not show the initial power law decay, yielding consistent exponents.

Generally, the Lyapunov exponent exhibits two distinct regimes. In the high-temperature, short-range correlated regime, there is little temperature dependence. This is not at all surprising: the state of a system with a bounded local energy spectrum changes only little when the temperature is raised well above this bandwidth.  {In this regime, $\lambda\sim |J|$, only weakly dependent on the dimension, is thus determined by the local microscopic physics dictated by the spin-exchange energy-scale $J$.}

The second, $T$-dependent regime, is entered
as correlations start to develop on a scale set by $J$, with a crossover to a well-developed low-temperature power-law regime,  $\lambda(T) \sim T^{\alpha}$ with $\alpha>0$. 
One may try to understand this decrease of the Lyapunov exponent at lower temperaturs from the reduction of the available phase space volume. As the energetic constraints become active around the cross-over temperature the entropy of the system is reduced, fluctuations decrease overall, and the dynamics slows down. Indeed, even for the cooperative paramagnet in $d=2$ for which the correlation length saturates to a small value in the low-$T$ limit, there is a power-law $\lambda(T) \sim T^{1/2}$ \cite{PhysRevLett.121.250602}.  Noticeably, in Fig. \ref{fig:lambda_vs_T}, while the high to low temperature crossover is smooth for $\lambda$ in $d=1,2$, the transition for $d=3$ leads to the appearance of a kink for both the FM and the AFM.

 The actual value of the exponent of the low-temperature scaling, $\alpha$, itself depends both on the dimension and the sign of the exchange constant $J$. Thus, unlike the thermodynamic static properties, including the position of the transition, which are equivalent for FM and AFM classically,   their {\it dynamic} chaotic behaviour and temperature scaling is different. This is of course not unexpected, in view of their differing respective (quadratic and linear) low energy dispersions. 

{Given the correlations are longer-ranged, and the long-wavelength quasiparticles increasingly well-developed (and strictly long-lived for $d=3$ below the transition), a slowing down of the dynamics compared to the cooperative paramagnet is again unsurprising: the power $\alpha$ in Fig. \ref{fig:lambda_vs_T} is greater than $1/2$ in all cases. However Eq.~\ref{eq:lyapunov} implies that chaos is dominated by the short spin-wave lifetimes, i.e.\  not from the vicinity of the Goldstone modes (Eq.~\ref{eq_fmhyd} and \ref{eq_afmhyd}). This, we believe, is also the reason why an evaluation of $\tau_{\bf k}$ for long-wavelengths fails to account for the observed value of $\alpha$ in Fig. \ref{fig:lambda_vs_T}.}

\subsection{Butterfly velocity}

\begin{figure}
    \includegraphics[scale=0.95]{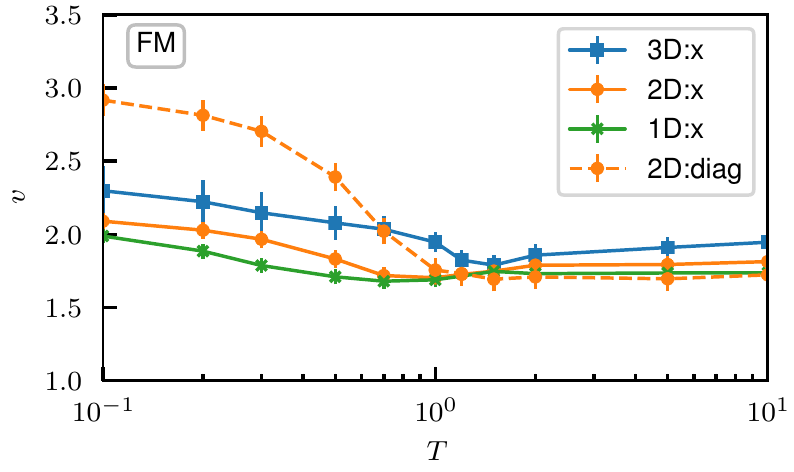}
    \includegraphics[scale=0.95]{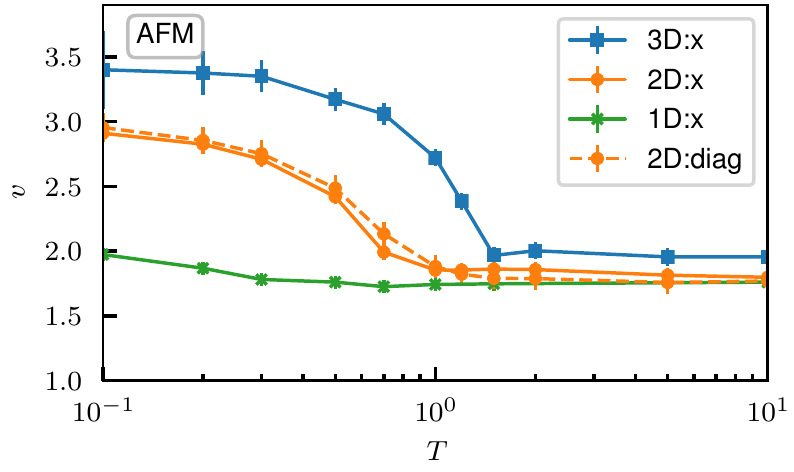}
    \caption{Temperature behaviour of the butterfly velocity for the FM (top) and the AFM (bottom) showing the low temperature correlated and the high temperature paramagnetic regime, separated by a cross-over ($d=1,2$) sharpening in the case of a true thermal phase transition for $d=3$. Results are obtained for systems with $L=10000$, $L=200$, and $L=40$ in $d=1,2,3$ respectively. Error bars are $1\sigma$ standard deviations of the fits. Solid lines are extracted from cuts along lattice-axes, e.g. $y=z=0$, dashed lines for 2D along the diagonal $x=y$.
    \label{fig:v_vs_T}
    }
\end{figure}

The temperature dependence of the butterfly velocity is shown in  Fig.~\ref{fig:v_vs_T}. This is obtained from fits to the position along lattice axes where the decorrelator exceeds a threshold $D_0$, e.g. $x_{thr} = v t$ with $D(x_{thr},t) > D_0$ for $D_0=10^{-10}$. The resulting velocity is independent of the threshold  for sufficiently small thresholds in the range we can probe, and consistent with the propagation velocity of the main peak (see Fig.~\ref{fig_tslicefm}). We however note that for the FM in the ordered regime it will be direction dependent, i.e. the speed will be minimal along lattice-axes, and maximal along the body-diagonal due to the cubic form of the wave-fronts.

In the high temperature regime, note that a light-cone arises despite the absence of spin waves and their `ballistic' propagation. In fact 
the dynamic spin-spin correlator in this regime is diffusive \cite{PhysRevLett.121.024101,PhysRevLett.121.250602}. As for the Lyapunov exponent, in this regime the butterfly speed is essentially constant and determined by the strength of the exchange coupling.

In contrast, in the  low-temperature regime, the spin-waves as shown above are well defined quasi-particles even deep into the chaotic regime. In particular the velocity calculated in the previous section from the free de-correlator agrees well with the butterfly velocity for the lowest measured temperatures, particularly for the antiferromagnet where the wavefront is isotropic and along the body-diagonal for the 
FM where we predict $v_{b} = 2 \sqrt{d}$, e.g. $2$, $2.82$ and $3.46$ in $d=1,2,3$.

For $d=3$ there is a gradual hardening of the butterfly velocity across the transition temperature. This is more prominent for the antiferromagnet, where it seems to follow the pronounced increase of the spin stiffness, which determines the spin-wave velocity via Eq.~\ref{eq_afmhyd}. 
Ref.~\cite{ruidas2020manybody} also studied the Lyapunov exponent and butterfly velocity across a BKT and Ising transition in two-dimensional XXZ spin models, with similar results as ours detailed above. Intriguingly, Ref.~\cite{ruidas2020manybody} did not detect any sharp signature of phase transitions in the these measures of chaos.

\section{Comparison between ordered magnet and spin liquid \label{sec:SL_vs_FM}}
The role of ordering with the concomitant appearance of quasiparticles can be crisply juxtaposed to the situation in the case of the classical kagome spin-liquid \cite{PhysRevLett.121.250602}, where no ordering occurs down to $T=0$ on account of the geometric frustration of the spin interactions.  Both the Lyapunov exponent and the light-cone velocity show smooth crossovers from the high-temperature paramagentic regime (common to both cases) to the low-temperature cooperative paramagnetic, or spin-liquid, regime. In this regime, both quantities vanish algebraically with temperature, obeying the following relation with the spin diffusion constant \cite{PhysRevLett.121.250602}:
\begin{equation}
D\sim v_B^2/\lambda \ .
\label{eq:balldiff}
\end{equation}

The coexistence of the {\it diffusion} constant, $D$, with the {\it ballistic} butterfly speed $v_B$ may at first sight seem somewhat surprising. {Below, we first} provide a simple picture for how these two different types of behaviour can coexist naturally. This yields a remarkable connection between the physics of chaos, and that of hydrodynamics, linking spatio-temporal chaos time- and length-scales to the transport coefficient of a conserved charge.

We then contrast  this expression with the situation in a system with quasiparticles, where the diffusion constant {within an  kinetic theory set-up appropriate for this situation, is given in terms of a characteristic velocity associated with the quasi-particles, $v_\mathrm{qp}$, and their scattering rate, $\lambda_\mathrm{qp}$, 
\begin{equation}
D\sim v_\mathrm{qp}^2/\lambda_\mathrm{qp} \ .
\label{eq:qpdiff}
\end{equation}}
Collecting our previous results implies that the butterfly velocity is simply replaced by the characteristic speed of ballistic propagation of the quasiparticles, with $\lambda_\mathrm{qp}$ related to the  Lyapunov exponent via Eq.~\ref{eq:lyapunov}. {Then, from our above discussion we conclude that while for antiferromagnet $v_\mathrm{qp}$ indeed represents the velocity of the long wavelength spin waves as calculated from the hydrodynamic theory, for the ferromagnet  modes away from the ordering modes set this scale.}

\subsection{Coexistence of diffusive and ballistic correlators}
For {an insight into the form} of Eq.~\ref{eq:balldiff}, it is convenient to define the average, and difference, combinations of the two copies, $ {\bf S}_T^{\mathrm{I}}, {\bf S}_T^{\mathrm{II}}$ introduced to define the decorrelator in Eq.~\ref{eq_decorrelator}. Both the total difference (Eq.~\ref{eq_dels}) 
\begin{align}
    {\bf S}_T^{-}=\sum_{i}\delta{\bf S}_i
    \label{eq_dels_2}
\end{align}
as well as the total average magnetisation
\begin{align}
    {\bf S}_T^{+}=\frac{1}{2}\sum_i\left({\bf S}^{\mathrm{I}}_i+{\bf S}^{\mathrm{II}}_i\right):=\sum_i{\bf S}_i^+
    \label{eq_syms}
\end{align}
are conserved. Therefore the corresponding densities of $\delta{\bf S}_i$ and ${\bf S}_i^+$ are expected to diffuse (it is straightforward to see from Eq.~\ref{eq_eom} that the dynamics of the two densities, though coupled with each other, are also local) with diffusion constants (say) $D_-$ and $D_+$ respectively. These diffusions can be studied through the respective two two-point correlators
\begin{align}
C_+({i},t)=\langle{\bf S}^+_{{i},t}\cdot{\bf S}^+_{{\bf 0},0}\rangle_T
\label{eq_cp}
\end{align}
and 
\begin{align}
C_-({i},t)=\langle\delta{\bf S}_{{i},t}\cdot\delta{\bf S}_{{\bf 0},0}\rangle_T
\label{eq_cm}
\end{align}
and we expect 
\begin{align}
C_\pm({i},t)\sim \frac{1}{\sqrt{D_\pm t}}e^{-{\bf x}_i^2/(D_\pm t)}
\end{align}

\begin{figure}
    \centering
    \includegraphics[width=0.24\textwidth]{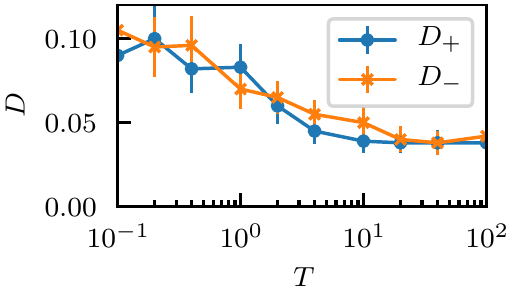}%
    \includegraphics[width=0.24\textwidth]{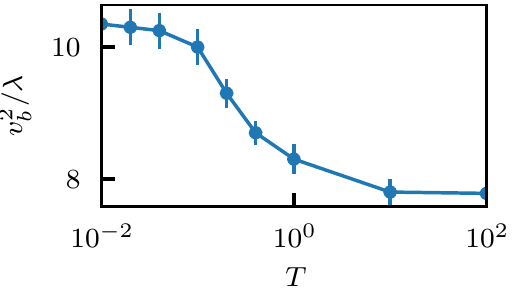}
    \caption{Left panel: Diffusion constants $D_{\pm}$  of the (anti) symmetric correlators versus temperature $T$. Diffusion constants are defined from the low $q$ dependence of the structure factor, $\mathcal{S}_{\pm}(q,\omega) =\sum_x \int_t e^{i \omega t} e^{-i \vect{q} \vect{x}} C_{\pm}(x,t)$, via $\mathcal{S}_{\pm}(q,\omega) \sim \frac{1}{ \omega^2 +\kappa(q)^2}$, with $\kappa(\vect{q}) = D q^2$ depending quadratically on $\vect{q}$ around $\vect{q}=0$. Right Panel: Ratio $v_b^2/\lambda$ of the square of the butterfly velocity to the Lyapunov exponent versus temperature $T$. The Lyapunov is defined from $D(x=0,t) \sim e^{2 \lambda t}$. \label{fig:diffusion}}
\end{figure}

Indeed for the kagome spin liquid, these two diffusion constants are plotted as a function of temperature in Fig. \ref{fig:diffusion} (left panel).

For our protocol, the diffusion of $\delta{\bf S}_{it}$ suggest that the difference field evolves as
\begin{align}
\delta{\bf S}_{{i},t}\sim\varepsilon{\bf n}_{{i},t}~e^{-{\bf x}_i^2/(2D_-t)}
\label{eq_deldiff}
\end{align}
where is ${\bf n}_{i,t}$ is a vector that encodes  {magnitude and direction of the} fast local fluctuations in the difference field. This describes the motion of the difference field throughout the system starting from $i=0$ at $t=0$ such that
\begin{align}
{\bf S}_{T}^-=\boldsymbol{\epsilon}
\label{eq_difftot}
\end{align}
{for all times.}

For a chaotic system, the growth of this {difference field at a given location, $i$,}  is expected to be exponential, $\exp(\lambda t)$, for arbitrarily long times in the limit $\epsilon\rightarrow0$. {While the random directions of ${\bf n}_{i,t}$ ensure that Eq.~\ref{eq_difftot} is obeyed, i.e.\ the decorrelator, is expected to have a form}
\begin{align}
\mathcal{D}({\bf x}_i,t)&=\varepsilon^2e^{\lambda t}e^{-\frac{{\bf x}_i^2}{D_-t}}=\varepsilon^2e^{\lambda t\left(1-\frac{{\bf x}_i^2}{v_B^2t^2}\right)}
\label{eq_deco}
\end{align}
where $\lambda$ is the Lyapunov exponent and by definition the butterfly speed, $v_B=\sqrt{D_-\lambda}$. 

The constants $D_\pm$ refer to the diffusion of two different conserved quantities, and they need not be equal in general. This is most prominent in case that the diffusion is due to separate set of objects (particles/excitations)  that carry one of the two charges of symmetric and staggered {\it spin-rotation symmetry}, $O_\pm(3)$. However, in case the same objects carry both the charges, then a Wiedemann-Franz type law can emerge, {but now} relating the two diffusion constants, 
\begin{align}
\frac{D_+}{D_-}=\eta \ ,
\end{align}
would be expected, where $\eta$ is a constant which may depend on the energy scales, as well as details, of the system under consideration.
In this case, the spin diffusion constant is
\begin{align}
    D_+\propto D_-\sim \frac{v_B^2}{\lambda}\ ,
    \label{eq_diff_chaos}
\end{align}
as advertised above, Eq.~\ref{eq:balldiff}. {The numerical verification of these ideas is shown in the right panel of Fig.~\ref{fig:diffusion} for the classical kagome spin liquid which exhibits no ordering at any $T$, thereby allowing fits over a large range of $T$. }

\subsection{Diffusion and chaos with and without quasiparticles}

We now return to the observation that a similar relationship  between the chaos and diffusion holds both for  the symmetry broken case, Eq.~\ref{eq:qpdiff},  as well as the spin liquid, Eq.~\ref{eq:balldiff}. 

In case of the ordered phase, we obtained a relation between the Lyapunov exponent and a time scale  $\lambda\sim \tau^{-1}$, Eq.~\ref{eq:lyapunov}, where $\tau$ is an appropriately defined single-particle lifetime of the low energy quasi-particles, the spin waves. These spin waves
propagate with a velocity $v_{\mathrm{qp}}\sim J$, which subsumes the ballistic propagation of the perturbation wave front, $v_B\sim v_{\mathrm{qp}}$. In this situation, the kinetic theory of dilute gases yields a diffusion constant, $D_+\sim v_{\mathrm{qp}}^2\tau\sim v_{\mathrm{B}}^2/\lambda$. 

Note that, despite the visual similarity between Eq.~\ref{eq:qpdiff} and Eq.~\ref{eq:balldiff}, their underlying physics differs fundamentally: there is no straightforward quasi-particle description for the spin liquid, whose low-energy sector with its huge ground state degeneracy is completely unlike that of the ordered magnet with its emergent integrability and long-lived quasiparticles. In this sense, the emergence of the butterfly velocity described in the previous subsection is an entirely separate, and remarkable, feature of many-body chaos at low temperature.

In passing, we note that for the symmetric combination, {\it i.e.}, the average magnetisation density, the fast local fluctuations, however cannot grow exponentially due to the already large background present in the form of a local spin-length. Therefore while the constraint (that can be derived from the equation of motion for the spins)
\begin{equation}
\partial_t({\bf S}^+_i\cdot{\bf S}^+_i+\delta{\bf S}_i\cdot\delta{\bf S}_i)=0
\end{equation}
indicates that the symmetric part contains the same information about the chaotic properties, e.g. the butterfly velocity and the Lyapunov, as the decorrelator, we note that the signal might in practice be hidden under the magnetic fluctuations and impossible to extract from the symmetric part.  However, as noted in Ref. \onlinecite{PhysRevE.89.012923}, in classical spin systems without spin conservation (say a XYZ model) the initial difference of total magnetisation between the two copies grows exponentially.

\section{Summary and Outlook\label{sec:discussion}}

In summary, we have presented a study of chaos and its temperature dependence in a family of model Hamiltonian many-body system which allows considerable variation in terms of choice of lattice, dimension and interaction. This provides access to different thermodynamic phases, such as a disordered [paramagnetic], ordered [(anti)ferromagnetic], as well as the critical  regime separating these. This has in turn permitted us to identify a number of distinct regimes characterised by different natural degrees of freedom, transport mechanisms, and concomitant velocity and time scales.

Our combined numerical and analytical investigations concretely connect the signatures of many-body chaos in classical spin systems,  the Lyapunov exponent $\lambda$ and the butterfly velocity $v_B$, with the velocity $v_\mathrm{qp}$ and scattering time $\tau_\mathrm{qp}$ associated with the spin waves, i.e.\ the quasiparticles in the phase which  spontaneously breaks spin rotation symmetry. These relations are directly manifested in the de-correlator, which quantifies how two copies with weakly, and locally (in real space), perturbed initial conditions diverge in time and space. 

At low $T$ and short times ($t< \lambda^{-1}$) integrable behaviour emerges whose butterfly velocity and  power-law temporal decay are concretely captured within linear spin-wave theory. Interestingly, the shape of the wave-front may depend on the interactions, being `hypercubic' in shape for the nearest neighbour ferromagnet in $d>1$, while the antiferromagnet exhibits isotropic `hyperspherical' behaviour. This is quite a tangential point to the present study, but besides some early work \cite{PhysRev.146.387}, we are not aware of a systematic study of the properties of freely propagating disturbances across lattices as a function of Hamiltonian parameters. Indeed it would be an interesting question how the symmetries of the considered lattices and further-neighbour couplings affect the observed wave-fronts. 
We also leave open the fate of the shape of these wave-fronts at longer times when interactions become relevant, which might be sufficient to restore isotropy even on these lattices.

The integrability, however, is only approximate and thus a transient phenomenon, even though the corresponding timescale, the spin-wave lifetime, $\tau_{\bf k}$ (where ${\bf k}$ is the spin-wave momentum),  grows as the temperature is decreased. The short time integrable behaviour thus gives way to fully developed chaos as witnessed by the exponential temporal growth of the decorrelator in addition to its ballistic spread through an intermediate scarred region. We interpret this a consequence of well defined quasi-particles, namely the spin-waves, transporting weight of the decorrelator ballistically while undergoing repeated scattering events. This results in random-walk like behaviour with diffusive core of the decorrelator on top of the exponential chaotic temporal growth. The novel scarred regime still holds plenty of interest for further study, for instance regarding the detailed mechanism leading to the generation of the secondary lightcones, as well as their statistical distribution in what appears to be a regime dominated by rare events. 

In the ordered low temperature regime, the Lyapunov exponent also vanishes as a power law in temperature, albeit with a different exponent which in turn depends on  both dimension and sign of the coupling; at the same time, the light-cone velocity  gets subsumed by the velocity of the ballistically propagating quasi-particles, and also saturates to a finite value at low temperatures.
We also find that both the Lyapunov and the light-cone velocity show characteristic features at the phase-transition, where the Lyapunov changes from esentially constant in the paramagnetic regime to a powerlaw decay, and the velocity shows a minimum  for the ferromagnet and a characteristic stiffening for the $d=3$ antiferromagnet reflecting the emergent spin stiffness.

We note that the same spin-wave scattering is responsible for the thermalisation of the {\it weakly interacting gas} of spin-waves. This notion of a dilute thermalised gas of spin-waves then forms the basis of the kinetic theory of transport at low temperature. Indeed the above picture is generic and forms one of the central pillars of low temperature transport theory in symmetry broken systems, and our work ties this in with the nature of the concomitant many-body chaos.

Such a low energy transport theory can be contrasted to the low temperature cooperative paramagnet, where quasiparticles are absent but diffusive transport persists. There, the transport coefficients are directly determined by the chaos timescales and lengthscales, with both Lyapunov exponent and  light-cone velocity exhibiting a power-law temperature dependence.

{The present observation of the importance of chaos time/length scales for hydrodynamics in presence and absence of quasiparticles quantitatively indicates an intriguing and important role of many-body chaos in the transport of strongly correlated systems which we think are of much broader significance in both classical and quantum many-body systems.} Indeed, the connection between classical and quantum chaos presents one of the most fascinating aspects of the field of many-body dynamics at present. While for the quantum setting a number of results -- under the headings of bounds for chaos and relatedly, Planckian transport -- have been established, their fate in the semiclassical limit is a subject of current investigation \cite{kurchan2016quantum,scaffidi2017semiclassical,yin2020quantum}.

We would like to end with two open interesting questions pertaining to two well-known frameworks of transport in the context of symmetry breaking and thermal phase transitions. The first pertains to the connection between the present observations and the elegant theory of non-linear fluctuating hydrodynamics \cite{PhysRevLett.111.230601} for the dynamic correlation function which characterises the long distance and time scaling of the low energy modes in the broken symmetry phase. Application of such approaches to classical spin systems in one dimensions are rather recent \cite{das2020nonlinear} and their connection to chaos remains to be understood. The second is related to the similar connection between our present approach and the theory of dynamic critical phenomena \cite{PhysRev.188.898}.  While the characterisation of chaos near a critical point has been recently addressed in model systems \cite{Schuckert_2019,ruidas2020manybody}, concrete connection to the different forms of coarse grained hydrodynamics and their possible relationship with many-body chaos remains open. In either case the systematic understanding of the mode-coupling there for the difference field, as developed here, and its similarity with the spin-wave dynamics can  serve as a starting point of such an understanding.

\begin{acknowledgements}
We acknowledge fruitful discussions with Sumilan Banerjee,  Abhishek Dhar, Benoit Dou\c{c}ot, Sean Hartnoll, Dima Kovrizhin, Anupam Kundu, Subroto Mukerjee, Samriddhi Sankar Ray, and Shivaji Sondhi. SB acknowledges funding from a Max Planck Partner Grant at ICTS and the hospitality of visitors program at MPIPKS; SERB-DST (India) for funding through project grant No. ECR/2017/000504 and the support of the Department of Atomic Energy, Government of India, under project no.12-R\&D-TFR-5.10-1100.
This work was in part supported by the Deutsche Forschungsgemeinschaft  under grants SFB 1143 (project-id 247310070) and the cluster of excellence ct.qmat (EXC 2147, project-id 390858490). 

\end{acknowledgements}

\begin{appendix}

\section{Spin wave analysis}
\label{appen_sw}

Here,  we provide details of the spin-wave analysis for the nearest neighbour classical ferromagnets and  antiferromagnets in $d=1,2,3$. From the equation of motion (Eq.~\ref{eq_eom}) we get, by using the form of the spins in Eq.~\ref{eq_spin_wave}, the dynamics of the spin-waves as
\begin{align}
\partial_t{\bf L}_i&=\left[{\bf n}_i[1-{\bf L}_i^2/2]+{\bf L}_i\right]\times\sum_j J_{ij} \left[{\bf n}_j[1-{\bf L}_j^2/2]+{\bf L}_j\right]
\label{eq_sw}
\end{align}
\subsection{Nearest Neighbour Ferromagnet}

For the ferromagnet,  $J<0$, the ground state is given by ${\bf n}_i=\hat{\bf z}$ (and $\hat{\bf n}\perp {\bf L}$). The transverse fluctuations are described by 
\begin{align}
\partial_t{\bf L}_i=&-|J|\hat{\bf z}\times\left[\sum_{j\in NN_i}{\bf L}_j-\sum_{j\in NN_i}{\bf L}_i\right]\nonumber\\
&+\frac{|J|}{2}{\bf z}\times\left[{\bf L}_i^2\sum_{j\in NN_i} {\bf L}_j-{\bf L}_i\sum_{j\in NN_i}~{\bf L}_j^2\right]
\end{align}
which in Fourier space becomes
\begin{align}
    \partial_t{\bf L}_{{\bf k}}=&\gamma({\bf k})\mathcal{Z}\cdot{\bf L}_{{\bf k}}+\frac{1}{N}\sum_{\bf q}\mathcal{M}_{\bf k,q}\cdot{\bf L}_{\bf k-q}
    \label{eq_spinwave_nonlinear}
\end{align}
where 
\begin{align}
{\bf L}_i=\frac{1}{N}\sum_{\bf k}e^{-i{\bf k\cdot r_i}}{\bf L}_{\bf k}
\label{eq_ftl}
\end{align}
\begin{align}
\mathcal{Z}=\left(\begin{array}{ccc}
0 & -1 & 0\\
1 & 0 & 0\\
0 & 0 & 0\\
\end{array}\right)
\label{eq_zmat}
\end{align}
and 
\begin{align}
    \mathcal{M}_{\bf k,q}=-\frac{(\gamma_{\bf k-q}-\gamma_{\bf q})}{2N}\sum_{\bf q'}{\bf L_{q'}}\cdot{\bf L}_{\bf q-q'}\mathcal{Z}
    \label{eq_fmmkq}
\end{align}
where $\gamma_{\bf k}$ is given by Eq.~\ref{eq_ferro_disp}. The linear solution is obtained by considering the bare Green's function 
\begin{align}
G_0(\omega,{\bf k})=\left[i\omega-\gamma({\bf k})\mathcal{Z}\right]^{-1}
\end{align}
or
\begin{align}
{G}_0({\bf k},t)
=&\left(\begin{array}{ccc}
\cos(|\gamma_{\bf k}| t) & -\sin(|\gamma_{\bf k}| t) & 0\\
\sin(|\gamma_{\bf k}| t) & \cos(|\gamma_{\bf k}| t) & 0\\
0 & 0 & 1\\
\end{array}\right)
\label{eq_free_propagator}
\end{align}
leading to
\begin{align}
{\bf L}_{{\bf q};t}=G_0({\bf q},t)\cdot {\bf L}_{{\bf q};0}
\end{align}
which, written explicitly (in real space), is of the form
\begin{align}
L^x_i&=\frac{1}{N}\sum_{\bf q}e^{-i{\bf q}\cdot{\bf r}_i}\left[A_{\bf q}e^{i|\gamma({\bf q})|t}+B_{\bf q}e^{-i|\gamma({\bf q})|t}\right]\\
L^y_i&=\frac{-i}{N}\sum_{\bf q}e^{-i{\bf q}\cdot{\bf r}_i}\left[A_{\bf q}e^{i|\gamma({\bf q})|t}-B_{\bf q}e^{-i|\gamma({\bf q})|t}\right]
\label{eq_fmspinwaves}
\end{align}
where
$$A_{\bf q}=\frac{L^x_{{\bf q};0}+iL^y_{{\bf q};0}}{2}~~~{\rm and}~~~~~~~B_{\bf q}=\frac{L^x_{{\bf q};0}-iL^y_{{\bf q};0}}{2}\ . $$

Therefore $\mathcal{M}_{\bf k,q}$ represents three spin-wave modes scattering, i.e., the leading scattering term.
Within a ${\bf k}$-dependent relaxation time approximation, we can re-write Eq.~\ref{eq_spinwave_nonlinear} as
\begin{align}
    \partial_t{\bf L}_{{\bf k}}=&\gamma({\bf k})\mathcal{Z}\cdot{\bf L}_{{\bf k}}-\frac{1}{\tau_{\bf k}}{\bf L}_{\bf k}
\end{align}
where $\tau_{\bf k}$ is the lifetime of a spin-wave. This is similar to the Landau-Lifshits-Gilbert equation.

\subsection{Nearest neighbour bipartite antiferromagnet (Neel order)}

For an N\'eel state (Eq.~\ref{eq_afmorder}), as in the ferromagnetic case, Eq.~\ref{eq_sw} gives

\begin{align}
\partial_t{\bf L}_{i,A}=&J\hat{\bf z}\times\left[2d {\bf L}_{i,A}+\sum_j  {\bf L}_{j,B}\right]\\
&-\frac{J}{2}\hat{\bf z}\times\sum_{j}\left[{\bf L}_{i,A}^2  {\bf L}_{j,B}+{\bf L}_{i,A}\sum_j {\bf L}_{j,B}^2\right]
\end{align}
and 
\begin{align}
\partial_t{\bf L}_{i,B}=&-J\hat{\bf z}\times\left[2d {\bf L}_{i,B}+\sum_j  {\bf L}_{j,A}\right]\\
&+\frac{J}{2}\hat{\bf z}\times\sum_{j}\left[{\bf L}_{i,B}^2  {\bf L}_{j,A}+{\bf L}_{i,B}\sum_j {\bf L}_{j,A}^2\right]\ .
\end{align}
Now, defining
\begin{align}
{\bf L}_\pm={\bf L}_A\pm {\bf L}_B
\end{align}
we get, after Fourier transforming,
\begin{align}
    \partial_t\left[\begin{array}{c}
    {\bf L}_{\bf k+}\\
    {\bf L}_{\bf k-}\\
    \end{array}\right]=&\mathcal{Z}\left[\begin{array}{cc}
    0 & \gamma_{\bf k}\\
    (4dJ-\gamma_{\bf k}) & 0\\
    \end{array}\right]\left[\begin{array}{c}
    {\bf L}_{\bf k+}\\
    {\bf L}_{\bf k-}\\
    \end{array}\right]\nonumber\\
    &+\frac{1}{N}\sum_{\bf q}\left[\begin{array}{cc}
    \mathcal{M}_{\bf k,q}^+ & -\mathcal{N}_{\bf k,q}^-\\
    -\mathcal{R}_{\bf k,q}^+ & \mathcal{M}_{\bf k,q}^-\\
    \end{array}\right]\left[\begin{array}{c}
    {\bf L}_{\bf k-q;+}\\
    {\bf L}_{\bf k-q;-}\\
    \end{array}\right]
    \label{eq_afm_sw}
\end{align}

\begin{align}
    \mathcal{M}^+_{\bf k,q}&=\frac{1}{4N}(\gamma_{\bf k-q}-\gamma_{\bf q})\sum_{\bf q_1}\left({\bf L_{\bf q_1}}^+\cdot{\bf L}^-_{\bf q-q_1}\right)\mathcal{Z}\\
    \mathcal{M}^-_{\bf k,q}&=\frac{1}{4N}(4Jd-\gamma_{\bf k-q}-\gamma_{\bf q})\sum_{\bf q_1}\left({\bf L_{\bf q_1}}^+\cdot{\bf L}^-_{\bf q-q_1}\right)\mathcal{Z}
    \label{eq_mpmafm}
\end{align}
\begin{align}
    \mathcal{N}_{\bf k,q}&=\frac{\gamma_{\bf k-q}-\gamma_{\bf q}}{8N}\sum_{\bf q_1}\left[{\bf L_{\bf q_1}}^+\cdot{\bf L}^+_{\bf q-q_1}+{\bf L_{\bf q_1}}^-\cdot{\bf L}^-_{\bf q-q_1}\right]\mathcal{Z}\\
    \mathcal{R}_{\bf k,q}&=\frac{\tilde\gamma_{\bf k-q}+\tilde\gamma_{\bf q}}{8N}\sum_{\bf q_1}\left[{\bf L_{\bf q_1}}^+\cdot{\bf L}^+_{\bf q-q_1}+{\bf L_{\bf q_1}}^-\cdot{\bf L}^-_{\bf q-q_1}\right]\mathcal{Z}
    \label{eq_nrafm}
\end{align}

From this, again similarly to the FM case, for the free spin-waves, we get
\begin{align}
\partial_t {\bf L}_{{\bf k}\pm}^0&=\mp(\tilde\gamma_{\bf k}\mp 2dJ)~\hat{\bf z}\times{\bf L}_{{\bf k}\mp}^0
\label{eq_swlin}
\end{align}
where 
\begin{align}
\tilde\gamma_{\bf k}=2J\sum_{i=1}^d\cos({\bf k\cdot\delta})=2dJ-\gamma_{\bf k}.
\label{eq_tgamma}
\end{align}
This gives
\begin{align}
\partial^2_t {\bf L}_{{\bf k}+}^0=-\Gamma_{\bf k}^2~{\bf L}_{{\bf k}+}^0\ ,
\end{align}
where 
\begin{align}
   \Gamma_{\bf k}=2J\left(d^2-(\tilde\gamma_{\bf k}/2J)^2\right)^{1/2}.
   \label{eq_bigamma}
\end{align}
Therefore we have
\begin{align}
{\bf L}_{{\bf k}+}^0&={\bf A_k} e^{i|\Gamma_{\bf k}|t}+{\bf B_k}e^{-i|\Gamma_{\bf k}|t}\\
{\bf L}_{{\bf k}-}^0&=-\frac{i}{\rho_{\bf k}}\left[{\bf C_k} e^{i|\Gamma_{\bf k}|t}-{\bf D_k}e^{-i|\Gamma_{\bf k}|t}\right]
\end{align}
where ${\bf C_k}=\hat{\bf z}\times {\bf A_k}$, ${\bf D_k}=\hat{\bf z}\times {\bf B_k}$ and
\begin{align}
    \rho_{\bf k}=\sqrt{\frac{d-\tilde\gamma_{\bf k}/2J}{d+\tilde\gamma_{\bf k}/2J}}.
    \label{eq_rhodef}
\end{align}

The above equations can be solved to get ${\bf A_k}$ and ${\bf B_k}$. From this we find that it is useful to define
\begin{align}
    {\bf U}_{{\bf k}+}^0&={\bf L}_{{\bf k}+}^0\\
    {\bf U}_{{\bf k}-}^0&=\rho_{\bf k}~\hat{\bf z}\times {\bf L}_{{\bf k}-}^0=\rho_{\bf k}\mathcal{Z}
   {\bf L}_{{\bf k}-}^0
    \label{eq_rotspinafm}
\end{align}
which diagonalizes the propagator in spin-space. Thus we get
\begin{align}
\left[\begin{array}{c}
     {\bf U}_{{\bf k}+}^0(t)\\
     {\bf U}_{{\bf k}-}^0(t)\\ 
\end{array}\right]=\left[\begin{array}{cc}
     g^0_{{\bf k}}& h^0_{{\bf k}}  \\
    -h^0_{{\bf k}} & g^0_{{\bf k}}\\
\end{array}\right]\left[\begin{array}{c}
     {\bf U}_{{\bf k};+}^0(0)  \\
     {\bf U}_{{\bf k};-}^0(0) \\ 
\end{array}\right]
\label{eq_matrixafmdecor}
\end{align}
where
\begin{align}
    g^0_{{\bf k}}=\left[\begin{array}{ccc}
    \cos(|\Gamma_{\bf k}|t) & 0 & 0\\
    0 & \cos(|\Gamma_{\bf k}|t) & 0\\
    0 & 0 & 1\\
    \end{array}\right]
\end{align}
and 
\begin{align}
    h^0_{{\bf k}}=\left[\begin{array}{ccc}
    \sin(|\Gamma_{\bf k}|t) & 0 & 0\\
    0 & \sin(|\Gamma_{\bf k}|t) & 0\\
    0 & 0 & 0\\
    \end{array}\right]
\end{align}

We re-write Eq.~\ref{eq_matrixafmdecor} as
\begin{align}
    \boldsymbol{ U}_{\bf k}^0(t)=\boldsymbol{\mathcal{G}}^{0}_{\bf k}(t)\cdot\boldsymbol{U}^0_{\bf k}(0)
    \label{eq_afm_free_propagator}
\end{align}

To understand the effect of scattering, it is useful to re-write Eq.~\ref{eq_afm_sw} as

\begin{align}
    \partial_t{\bf U}_{\bf k}=\boldsymbol{\chi_{\bf k}}\cdot\boldsymbol{U}_{\bf k}+\frac{1}{N}\sum_{\bf q}{\boldsymbol{\Xi}}^1_{\bf k,q}\cdot {\bf U}_{\bf k-q}
\end{align}
where
\begin{align}
    {\boldsymbol{\chi}}_{\bf k}=\left[\begin{array}{cc}
    0 & \Gamma_{\bf k}\\
    -\Gamma_{\bf k} & 0\\
    \end{array}\right]
    \label{eq_chieq}
\end{align}
describes the free evolution and 
\begin{align}
    {\boldsymbol{\Xi}}^1_{\bf k,q}=\left[\begin{array}{cc}
    \bar{\mathcal{M}}_{\bf k,q}^+ & \bar{\mathcal{N}}_{\bf k,q}^-\\
    \bar{\mathcal{R}}_{\bf k,q}^+ & \bar{\mathcal{M}}_{\bf k,q}^-\\
    \end{array}\right]
    \label{eq_xi1eq}
\end{align}
is the scattering matrix with elements given by
\begin{align}
    \bar{\mathcal{M}}^+_{\bf k,q}&=-\frac{1}{4N}(\gamma_{\bf k-q}-\gamma_{\bf q})\sum_{\bf q_1}\left[\frac{\boldsymbol{U_{\bf q_1+}}\cdot\boldsymbol{U}_{\bf q-q_1;-}}{\rho_{\bf q-q_1}}\right]\mathcal{Z}\\
    \bar{\mathcal{M}}^-_{\bf k,q}&=\frac{\rho_{\bf k}(\tilde\gamma_{\bf k-q}+\tilde\gamma_{\bf q})}{4N\rho_{\bf k-q}}\sum_{\bf q_1}\left[\frac{\boldsymbol{U_{\bf q_1+}}\cdot\boldsymbol{U}_{\bf q-q_1;-}}{\rho_{\bf q-q_1}}\right]\mathcal{Z}
    \label{eq_mpmafm_p}
\end{align}
\begin{align}
    \bar{\mathcal{N}}_{\bf k,q}&=-\frac{\gamma_{\bf k-q}-\gamma_{\bf q}}{8N\rho_{\bf k-q}}\sum_{\bf q_1}\left[\boldsymbol{U_{\bf q_1+}}\cdot\boldsymbol{U}_{\bf q-q_1;+}\right.\nonumber\\
    &~~~~~~~~~~~~~~~~~~~~~~~~~~~~~~~~~~~~~~~~\left.+\frac{\boldsymbol{U_{\bf q_1-}}\cdot\boldsymbol{U}_{\bf q-q_1;-}}{\rho_{\bf q_1}\rho_{\bf q-q_1}}\right]\\
    \bar{\mathcal{R}}_{\bf k,q}&=\frac{\rho_{\bf k}(\tilde\gamma_{\bf k-q}+\tilde\gamma_{\bf q})}{8N}\sum_{\bf q_1}\left[\boldsymbol{U_{\bf q_1+}}\cdot\boldsymbol{U}_{\bf q-q_1;+}\right.\nonumber\\
    &~~~~~~~~~~~~~~~~~~~~~~~~~~~~~~~~~~~~~~~~\left.+\frac{\boldsymbol{U_{\bf q_1-}}\cdot\boldsymbol{U}_{\bf q-q_1;-}}{\rho_{\bf q_1}\rho_{\bf q-q_1}}\right]
    \label{eq_nrafm_p}
\end{align}

This describes the scattering of the spin waves and leads to their finite lifetime.

\section{Decorrelator for the Ferromagnet}
\label{appen_ferrodecor}

From Eq.~\ref{eq_ds_sw_2}, for the ferromagnet with ${\bf n}_i=\hat{\bf z}$, the equation of motion for the difference field reads
\begin{align}
    \delta_t\delta{\bf S}_i=&-|J|\hat{\bf z}\times\left[\sum_{j\in NN_i}\delta{\bf S}_j-\sum_{j\in NN_i}\delta{\bf S}_i\right]\nonumber\\
&-|J|\left[{\bf L}_i\times\sum_{j\in NN_i}\delta{\bf S}_j+\delta{\bf S}_i\times\sum_{j\in NN_i}{\bf L}_j\right]\nonumber\\
&+\frac{|J|}{2}{\bf z}\times\left[{\bf L}_i^2\sum_{j\in NN_i} \delta{\bf S}_j-\delta{\bf S}_i\sum_{j\in NN_i}~{\bf L}_j^2\right]\ .
\end{align}

The Fourier transformation (Eq.~\ref{eq_fts}) of the above equation leads to (with ${\bf L}_{\bf k}=(L_{\bf k}^x\hat{\bf x}+L_{\bf k}^y\hat{\bf y})$) Eq.~\ref{eq_eom_mct_fm}, with in Eq.~\ref{eq_fmvertex},
\begin{align}
    \mathcal{O}_{\bf k,q}=(\gamma_{\bf k-q}-\gamma_{\bf q})\mathcal{L}_{\bf q}
    \label{eq_fmo}
\end{align}
where
\begin{align}
\mathcal{L}_{\bf q}(t)=\left(\begin{array}{ccc}
0 & 0 & L^y_{\bf q}\\
0 & 0 & -L^x_{\bf q}\\
-L^y_{\bf q} & L^x_{\bf q} & 0\\
 \end{array}\right)
\label{eq_lfm}
\end{align}
and $\mathcal{M}_{\bf k,q}$ is given by Eq.~\ref{eq_fmmkq}. The decorrelator can then be calculated from Eq.~\ref{eq_decorexp}. 
\subsection{Free solution}

\begin{figure}
\includegraphics{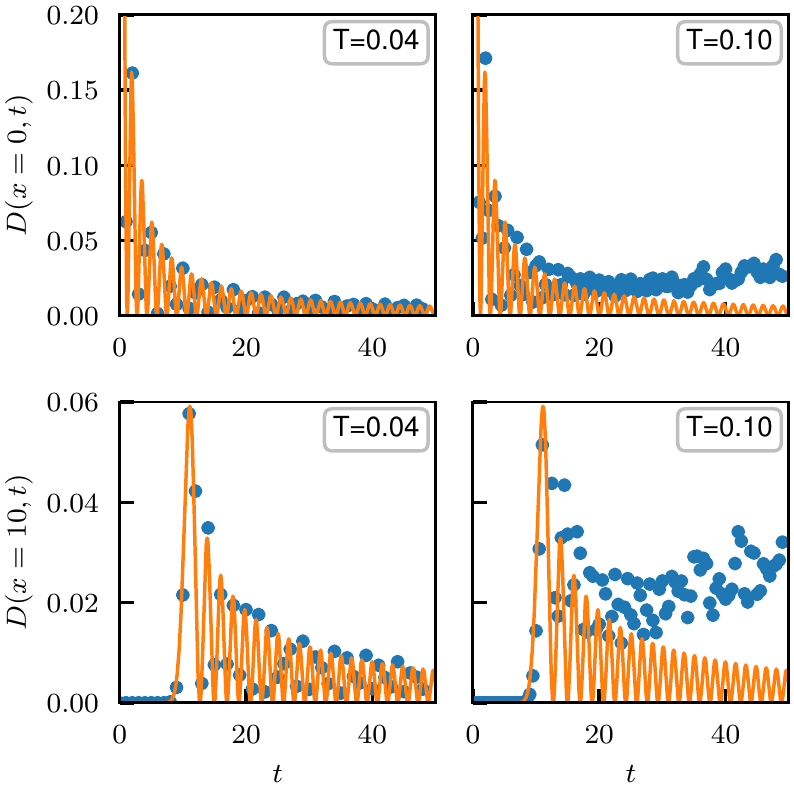}
\caption{{\bf Short time $t$ behaviour of $D(x,t)$ at fixed $x$ for the 1D ferromagnet} : Full numerical simulations (points) versus results of LSW (solid lines), Eq.~\ref{eq_zerodit_22_main}. Top panels at $x=0$, bottom panels at $x=10$. Temperature for the simulations are $T=0.04$ (left panels) and $T=0.10$ (right panels).
\label{fig_xslice_appen}}
\end{figure}

Neglecting scattering, the explicit form of the linear equation (first term on the right and side of Eq.~\ref{eq_eom_mct_fm}) of motion of $\delta{\bf S}_i$ is given by
\begin{align}
\partial_t\delta S^x_{\bf k}&=-\gamma_{\bf k}\delta S^y_{\bf k}\\
\partial_t\delta S^y_{\bf k}&=\gamma_{\bf k}\delta S^x_{\bf k}
\end{align}
whose solution is summarised in Eq.~\ref{eq_luv}. The initial conditions are obtained as follows. From Eqs. \ref{eq_dels} and \ref{eq_delsn}, we have Eq.~\ref{eq_sk_flat} such that
\begin{align}
\boldsymbol{\epsilon}=\frac{\varepsilon\left(-\sqrt{1-({\bf L}_0(0))^2}(L_0^x(0)\hat{\bf x}+L_0^y(0)\hat{\bf y})+({\bf L}_0(0))^2\hat{\bf z}\right)}{|{\bf L}_0(0)|}
\label{eq_initial_cond}
\end{align}

From this it follows that  
\begin{align}
    \eta=\sqrt{(\epsilon^x)^2+(\epsilon^y)^2}=\varepsilon \sqrt{1-|{\bf L}_0(0)|^2} \\
   \phi=\tan^{-1}(L^y(0)/L^x_0(0)).
\end{align}

Therefore, using Eq.~\ref{eq_deltas_SW}, the decorrelator reads 
\begin{widetext}
\begin{align}
\frac{\mathcal{D}(i,t)}{\varepsilon^2}&=(1-m_T^2)^2\delta_{i,0}+\frac{m_T^2}{(2\pi)^{2d}}\int_{BZ} d^d{\bf k}\int_{BZ}d^d{\bf k'}~\cos[(|\gamma[({\bf k})|-|\gamma({\bf k'})|)t+({\bf k+k'})\cdot{\bf r}_i]
\label{eq_zerodit_22_main_2}
\end{align}
\end{widetext}
which can be re-written as Eq.~\ref{eq_zerodit_22_main}  by using
\begin{align}
    J_\nu(t)=\frac{1}{2\pi i^\nu}\int_{-\pi}^\pi dk~e^{it\cos k}e^{i\nu k}
\end{align}
and associated properties of the Bessel function of the first kind, $J_\nu(x)$. In Eq.~\ref{eq_zerodit_22_main_2} we have also used the relation in thermodynamic equilibrium, $|{\bf L}_0(0)|^2=1-m_T^2$.


\subsection{Mode coupling and scattering of difference field with spin waves}

The mode-coupling expansion for $\delta{\bf S}_{\bf k}(t)$ is obtained by iterating Eq.~\ref{eq_mctfm} and is given by
\begin{widetext}
\begin{align}
\delta {\bf S}_{\bf k}(t)=&G_0({\bf k},t)\left[1+\frac{1}{N}\sum_{\bf q_1}\int_0^tdt_1~\mathcal{A}_{\bf k,q_1}(t_1)G_0({\bf k-q_1},t_1)\right.\nonumber\\
&~~~~~~~~~~~~~~+\frac{1}{N^2}\sum_{\bf q_1,q_2}\int_0^tdt_1~\mathcal{A}_{\bf k,q_1}(t_1)G_0({\bf k-q_1},t_1)\int_0^{t_1}dt_2~\mathcal{A}_{\bf k-q_1,q_2}(t_2)G_0({\bf k-q_1-q_2},t_2)\left.+\cdots\right]\cdot\boldsymbol{\epsilon}
\label{eq_feynman_mct}
\end{align}
\end{widetext}

From Eq.~\ref{eq_feynman_mct}, the expansion of the decorrelator is obtained via an expansion of $\delta {\bf S}_{\bf k}(t)\cdot\delta{\bf S}_{\bf k'}(t)$ whose perturbation series is given by \ref{eq_mctsummedfm}. Note that in deriving Eq.~\ref{eq_mctsummedfm} we have used 
\begin{align}
    \left[G^0_{\bf k}(t)\right]^T\cdot G^0_{\bf -k}(t)=1 \ .
\end{align}

\section{Decorrelator for the bipartite antiferromagnet}
\label{appen_afmdecor}

Next, we discuss the spread of decorrelations for the bipartite antiferromagnet. From Eq. \ref{eq_ds_sw_2}, using Eq.~\ref{eq_afmorder} for a bipartite antiferromagnet, we get
\begin{align}
\partial_t{\bf S}_{i,A}=&J\hat{\bf z}\times\left[2d~\delta{\bf S}_{i,A}+\sum_{j\in i}\delta{\bf S}_{j,B}\right]\nonumber\\
&+J\sum_{j\in i}\left[\delta{\bf S}_{i,A}\times{\bf L}_{j,B}+{\bf L}_{i,A}\times\delta{\bf S}_{j,B}\right]\nonumber\\
&-\frac{J}{2}\hat{\bf z}\times\sum_{j\in i}\left[{\bf L}_{i,A}^2\delta{\bf S}_{j,B}+\delta{\bf S}_{i,A}{\bf L}_{j,B}^2\right]
\end{align}
for sublattice $A$ and 
\begin{align}
\partial_t\delta{\bf S}_{i,B}=&-J\hat{\bf z}\times\left[2d~\delta{\bf S}_{i,B}+\sum_{j\in i}\delta{\bf S}_{j,A}\right]\nonumber\\
&+J\sum_{j\in i}\left[\delta{\bf S}_{i,A}\times{\bf L}_{j,B}+{\bf L}_{i,A}\times\delta{\bf S}_{j,B}\right]\nonumber\\
&+\frac{J}{2}\hat{\bf z}\times\sum_{j\in i}\left[{\bf L}_{i,B}^2\delta{\bf S}_{j,A}+\delta{\bf S}_{i,B}{\bf L}_{j,A}^2\right]
\end{align}
for sublattice $B$. Introducing the symmetric and antisymmetric modes for the difference field
\begin{align}
    \delta {\bf S}_{{\bf k}\pm}=\delta {\bf S}_{{\bf k},A}\pm\delta{\bf S}_{{\bf k},B}
\end{align}
similar to the antiferromagnetic spin-waves in the section above, Fourier transforming yields 
\begin{align}
    \partial_t\left[\begin{array}{c}
    \delta{\bf S}_{{\bf k}+}\\
    \delta{\bf S}_{{\bf k}-}\\
    \end{array}\right]=&\mathcal{Z}\left[\begin{array}{cc}
    0 & \gamma_{\bf k}\\
    (4dJ-\gamma_{\bf k}) & 0\\
    \end{array}\right]\left[\begin{array}{c}
    \delta{\bf S}_{{\bf k}+}\\
    \delta{\bf S}_{{\bf k}-}\\
    \end{array}\right]\nonumber\\
    &+\frac{1}{N}\sum_{\bf q}\left[\begin{array}{cc}
    \mathcal{A}_{\bf k,q}^+ & \mathcal{A}_{\bf k,q}^-\\
    \mathcal{B}_{\bf k,q}^+ & \mathcal{B}_{\bf k,q}^-\\
    \end{array}\right]\left[\begin{array}{c}
    \delta{\bf S}_{{\bf k-q};+}\\
    \delta{\bf S}_{{\bf k-q};-}\\
    \end{array}\right]\ .
    \label{eq_eom_afm_decor}
\end{align}

The various scattering terms are given by
\begin{align}
    \mathcal{A}^+_{\bf k,q}=&\frac{1}{2}\mathcal{O}_{\bf k,q}^++\mathcal{M}^+_{\bf k,q}\\
    \mathcal{A}^-_{\bf k,q}=&-\frac{1}{2}\mathcal{O}^-_{\bf k,q}-\mathcal{N}_{\bf k,q}\\
    \mathcal{B}_{\bf k,q}^+=&\frac{1}{2}\mathcal{P}^-_{\bf k,q}-\mathcal{R}_{\bf k,q}\\
    \mathcal{B}_{\bf k,q}^-=&\frac{1}{2}\mathcal{P}^+_{\bf k,q}+\mathcal{M}^-_{\bf k,q}
\end{align}
where
\begin{align}
    \mathcal{O}_{\bf k,q}^\pm=&(\gamma_{\bf k-q}-\gamma_{q})\mathcal{L}_{\bf q}^\pm\\
    \mathcal{P}^\pm_{\bf k,q}=&\pm\left(\tilde\gamma_{\bf q}+\tilde\gamma_{\bf k-q}\right)\mathcal{L}_{\bf q}^\pm
\end{align}
with $\mathcal{L}_{\bf q}^\pm$ being given by equations similar to Eq.~\ref{eq_lfm} with ${\bf L}_{\bf q}^\pm$ appropriately used for the two spin wave modes (see above); and $\mathcal{M}^\pm$, $\mathcal{N}$ and $\mathcal{R}$ are given by Eqs. \ref{eq_mpmafm} and \ref{eq_nrafm}.

\subsection{The free de-correlator}
Neglecting the scattering terms in Eq.~\ref{eq_eom_afm_decor}, the equation of motion for the free decorrelator is
\begin{align}
    \partial_t\left[\begin{array}{c}
    \delta{\bf S}_{{\bf k}+}^0\\
    \delta{\bf S}_{{\bf k}-}^0\\
    \end{array}\right]=&\mathcal{Z}\left[\begin{array}{cc}
    0 & \gamma_{\bf k}\\
    (4dJ-\gamma_{\bf k}) & 0\\
    \end{array}\right]\left[\begin{array}{c}
    \delta{\bf S}_{{\bf k}+}^0\\
    \delta{\bf S}_{{\bf k}-}^0\\
    \end{array}\right]
    \label{eq_free_eom_afm}
\end{align}

 Again due to the structure of the matrix $\mathcal{Z}$ (see Eq.~\ref{eq_zmat}), the longitudinal component (along the ordering direction) does not evolve. 
Explicitly solving Eq.~\ref{eq_free_eom_afm} for the transverse components we get
\begin{align}
\delta {\bf S}_{\bf K+}^0(t)=\epsilon^z\hat{\bf z}+\boldsymbol{\epsilon}_\perp \cos(|\Gamma_{\bf k}|t)+\hat{\bf z}\times\boldsymbol{\epsilon}_\perp\sin(|\Gamma_{\bf k}|t)~\rho_{\bf k} \label{eq:app_AFM_SKp}\\
\delta {\bf S}_{\bf K-}^0(t)=\epsilon^z\hat{\bf z}+\boldsymbol{\epsilon}_\perp \cos(|\Gamma_{\bf k}|t)+\hat{\bf z}\times\boldsymbol{\epsilon}_\perp\sin(|\Gamma_{\bf k}|t)~\frac{1}{\rho_{\bf k}} \label{eq:app_AFM_SKm}
\end{align}
where $\Gamma_{\bf k}$ and $\rho_{\bf k}$ are defined in Eqs. \ref{eq_bigamma} and \ref{eq_rhodef} respectively. Now, similarly to Eq.~\ref{eq_rotspinafm}, we define $\boldsymbol{\Delta}_{\bf k}$ as

\begin{align}
    \boldsymbol{\Delta}_{\bf k}\equiv\left[\begin{array}{c}
    \boldsymbol{\Delta}_{{\bf k};+}\\
    \boldsymbol{\Delta}_{{\bf k};-}\\
    \end{array}\right]=\left[\begin{array}{c}
    \delta{\bf S}_{{\bf k};+}\\
    \rho_{\bf k}~\hat{\bf z}\times\delta{\bf S}_{{\bf k};-}\\
    \end{array}\right]
    \label{eq_afmdeltadef}
\end{align}
with $\tilde\gamma_{\bf k}$ and $\rho_{\bf k}$ being defined by Eqs. \ref{eq_tgamma} and \ref{eq_rhodef}. The free solution can now be written as
\begin{align}
    \boldsymbol{\Delta}_{\bf k}^{0}(t)=\boldsymbol{\mathcal{G}}^0_{\bf k}(t)\cdot\boldsymbol{\Delta}_{\bf k}^0(0)
    \label{eq_freeafmdeltas}
\end{align}
where $\boldsymbol{\mathcal{G}^0_{\bf k}}$ is the free propagator defined in Eq.~\ref{eq_afm_free_propagator} and
\begin{align}
    \boldsymbol{\Delta}_{\bf k}(0)=\boldsymbol{\mathcal{V}}_{\bf k}\cdot\bar{\boldsymbol{\epsilon}}
\end{align}
where
\begin{align}
    \boldsymbol{\mathcal{V}}_{\bf k}=\left[\begin{array}{cc}
        1 & 0 \\
        0 &\rho_{\bf k} \mathcal{Z}\\
    \end{array}\right],~~~~~
    \bar{\boldsymbol{\epsilon}}=\left[\begin{array}{c}
         \boldsymbol{\epsilon}\\
         \boldsymbol{\epsilon}\\
    \end{array}\right]\ .
\end{align}

We note that in case of the antiferromagnet, we have used the initial conditions as given by Eq.~\ref{eq_sk_flat} and \ref{eq_initial_cond}. This breaks sublattice symmetry by having the perturbation initially at sublattice $A$. 

For sublattice A and B the free correlators are then given by
\begin{widetext}

\begin{align}
    \frac{\mathcal{D}^0_A(i,t)}{\varepsilon^2}=&(1-n_T^2)\delta_{i,0}+\frac{n_T^2}{2}\int_{BZ}~\frac{d^d{\bf k}}{(2\pi)^d}~\frac{d^d{\bf k'}}{(2\pi)^d}\left[\left\{1-\frac{1}{4}\left[\rho_{\bf k}+\frac{1}{\rho_{\bf k}}\right]\left[\rho_{\bf k'}+\frac{1}{\rho_{\bf k'}}\right]\right\}\cos[(|\Gamma_{\bf k}|+|\Gamma_{\bf k'}|)t]+{(\bf k+\bf k')\cdot r_i}]\right]\nonumber\\
    &+\frac{n_T^2}{2}\int_{BZ}~\frac{d^d{\bf k}}{(2\pi)^d}~\frac{d^d{\bf k'}}{(2\pi)^d}\left[\left\{1+\frac{1}{4}\left[\rho_{\bf k}+\frac{1}{\rho_{\bf k}}\right]\left[\rho_{\bf k'}+\frac{1}{\rho_{\bf k'}}\right]\right\}\cos[(|\Gamma_{\bf k}|-|\Gamma_{\bf k'}|)t]+{(\bf k+\bf k')\cdot r_i}]\right]
    \label{eq_afmdecor_freea}
\end{align}
and
\begin{align}
    \frac{\mathcal{D}^0_B(i,t)}{\varepsilon^2}=&\frac{n_T^2}{8}\int_{BZ}~\frac{d^d{\bf k}}{(2\pi)^d}~\frac{d^d{\bf k'}}{(2\pi)^d}~\left[\cos[(|\Gamma_{\bf k}|-|\Gamma_{\bf k'}|)t+{(\bf k+\bf k')\cdot r_i}]\left[\rho_{\bf k}-\frac{1}{\rho_{\bf k}}\right]\left[\rho_{\bf k'}-\frac{1}{\rho_{\bf k'}}\right]\right]\nonumber\\
    &-\frac{n_T^2}{8}\int_{BZ}~\frac{d^d{\bf k}}{(2\pi)^d}~\frac{d^d{\bf k'}}{(2\pi)^d}\left[\cos[(|\Gamma_{\bf k}|+|\Gamma_{\bf k'}|)t+{(\bf k+\bf k')\cdot r_i}]\left[\rho_{\bf k}-\frac{1}{\rho_{\bf k}}\right]\left[\rho_{\bf k'}-\frac{1}{\rho_{\bf k'}}\right]\right]
    \label{eq_afmdecor_freeb}
\end{align}
\end{widetext}
We note that the initial perturbation was put on sublattice A, as is evident from the above expressions and $n_T$ is the Neel order parameter. 

\subsection{Scattering and  chaos}

Incorporating the scattering, the equation of motion in terms of the $\boldsymbol{\Delta}_{\bf k}$ introduced in Eq.~\ref{eq_afmdeltadef} can be written as 

\begin{align}
    \partial_t\boldsymbol{\Delta}_{\bf k}=\boldsymbol{\chi}_{\bf k}\cdot\boldsymbol{\Delta}_{\bf k}+\frac{1}{N}\sum_{\bf q}\boldsymbol{\Xi}_{\bf k,q}\cdot\boldsymbol{\Delta}_{\bf k-q}
\end{align}
where $\boldsymbol{\chi}_{\bf k}$, given by Eq. \ref{eq_chieq}, controls the free evolution and $\boldsymbol{\Xi}_{\bf k,q}$ is the scattering kernel (that couples the different modes, similar to Eq.~\ref{eq_eom_mct_fm}) given by
\begin{align}
    {\boldsymbol{\Xi}}_{\bf k,q}={\boldsymbol{\Xi}}_{\bf k,q}^1+{\boldsymbol{\Xi}}_{\bf k,q}^2
\end{align}
with ${\boldsymbol{\Xi}}_{\bf k,q}^1$ given by Eq. \ref{eq_xi1eq} and
\begin{align}
  {\boldsymbol{\Xi}}_{\bf k,q}^2= \frac{1}{2}\left[\begin{array}{cc}
    \bar{\mathcal{O}}_{\bf k,q}^+ & \bar{\mathcal{O}}^-_{\bf k,q}\\
   \bar{\mathcal{P}}^-_{\bf k,q} & 0\\
    \end{array}\right]
\end{align}

with
\begin{align}
    \bar{\mathcal{O}}_{\bf k,q}^+=&(\gamma_{\bf k-q}-\gamma_{q})\bar{\mathcal{L}}_{\bf q}^+\\
    \bar{\mathcal{O}}_{\bf k,q}^-=&\frac{1}{\rho_{\bf q}\rho_{\bf k-q}}(\gamma_{\bf k-q}-\gamma_{q})\bar{\mathcal{L}}_{\bf q}^-\\
    \bar{\mathcal{P}}^-_{\bf k,q}=&-\frac{\rho_{\bf k}}{\rho_{\bf q}}\left(\tilde\gamma_{\bf q}+\tilde\gamma_{\bf k-q}\right)\bar{\mathcal{L}}_{\bf q}^-
\end{align}
where now we have
\begin{align}
\bar{\mathcal{L}}_{\bf q}^+(t)=\left(\begin{array}{ccc}
0 & 0 & U^y_{\bf q+}\\
0 & 0 & -U^x_{\bf q+}\\
-U^y_{\bf q+} & U^x_{\bf q+} & 0\\
 \end{array}\right)
\label{eq_lafmp}
\end{align}
and
\begin{align}
\bar{\mathcal{L}}_{\bf q}^-(t)=\left(\begin{array}{ccc}
0 & 0 & 0\\
0 & 0 & 0\\
U^y_{\bf q-} & -U^x_{\bf q-} & 0\\
 \end{array}\right)\ .
\label{eq_lafmm}
\end{align}
$\bar{\mathcal{M}}^\pm_{\bf k,q}$, $\bar{\mathcal{N}}_{\bf k,q}$ and $\bar{\mathcal{R}}_{\bf k,q}$ are given by Eqs. \ref{eq_mpmafm_p} and \ref{eq_nrafm_p}.

The calculation for $\boldsymbol{\Delta}_{\bf k}$ now proceeds very similarly as in the ferromagnet, leading to the solution, cf.\ Eq. \ref{eq_mctfm} :
\begin{align}
    \boldsymbol{\Delta}_{\bf k}(t)=\boldsymbol{\Delta}^0_{\bf k}(0)+\boldsymbol{\mathcal{G}}^0_{\bf k}(t)\cdot\sum_{\bf q}\int_0^t dt'~\boldsymbol{\Xi}_{\bf k,q}(t')\cdot\boldsymbol{\Delta}_{\bf k-q}(t')
    \label{eq_mctafm}
\end{align}
where the free solution $\boldsymbol{\Delta}^0_{\bf k}(t)$ is given by Eq.~\ref{eq_freeafmdeltas}. This can be expanded in terms of the free solution as
\begin{widetext}
\begin{align}
\boldsymbol{\Delta}_{\bf k}(t)=&\boldsymbol{\mathcal{G}}^0_{\bf k}(t)\left[\boldsymbol{\mathcal{V}}_{\bf k}+\frac{1}{N}\sum_{\bf q_1}\int_0^tdt_1~\boldsymbol{\Xi}_{\bf k,q_1}(t_1)\cdot\boldsymbol{\mathcal{G}}^0_{\bf k-q_1}(t_1)\cdot\boldsymbol{\mathcal{V}}_{\bf k-q}\right.\nonumber\\
&~~~~~~~~~~~~~~+\frac{1}{N^2}\sum_{\bf q_1,q_2}\int_0^tdt_1~\boldsymbol{\Xi}_{\bf k,q_1}(t_1)\cdot\boldsymbol{\mathcal{G}}^0_{\bf k-q_1}(t_1)\cdot\boldsymbol{\mathcal{V}}_{\bf k-q_1}\cdot\int_0^{t_1}dt_2~\boldsymbol{\Xi}_{\bf k-q_1,q_2}(t_2)\cdot\boldsymbol{\mathcal{G}}^0_{\bf k-q_1-q_2}(t_2)\left.+\cdots\right]\cdot\bar{\boldsymbol{\epsilon}}
\label{eq_feynman_mct_afm}
\end{align}
\end{widetext}
A diagrammatic representation similar to Eq.~\ref{eq_feynman_mct} is obtained for both $\delta{\bf S}$ as well as the decorrelator (not shown).  

In particular the summed de-correlator (Eq.~\ref{eq_sumdecorexp}) of the antiferromagnet is given by
\begin{align}
    \mathcal{I}(t)
    =\frac{1}{2}\sum_{\bf k}\langle\boldsymbol{\Delta}_{\bf k}\cdot\mathcal{H}_{\bf k}\cdot\boldsymbol{\Delta}_{\bf - k}\rangle_T
    \label{eq_summed_dec_afm}
\end{align}

where 
\begin{align}
    \mathcal{H}_{\bf k}=\left[\begin{array}{cc}
    1 & 0\\
    0 & \frac{1}{\rho_{\bf k}^2}\\
    \end{array}\right]
\end{align}

Therefore
\begin{widetext}
\begin{align}
    \frac{\boldsymbol{\Delta}_{\bf k}\cdot\mathcal{H}_{\bf k}\cdot\boldsymbol{\Delta}_{\bf -k}}{2}=&\epsilon^2+\frac{\bar{\boldsymbol{\epsilon}}^T}{2N}\cdot\sum_{\bf q}\int_0^t~dt_1\left[\boldsymbol{\mathcal{V}}_{\bf k}^T\cdot\mathcal{H}_{\bf k}\cdot\boldsymbol{\Xi}_{\bf k,q}(t_1)\cdot\boldsymbol{\mathcal{G}}^0_{\bf k-q}(t_1)\cdot\boldsymbol{\mathcal{V}}_{\bf k-q}\right.\nonumber\\
    &\quad\quad\quad\quad\quad\quad\quad\quad\quad\quad\quad\quad\left.+\left[\boldsymbol{\mathcal{V}}_{\bf k-q}\right]^T\cdot\left[\boldsymbol{\mathcal{G}}^0_{\bf k-q}\right]^T\cdot\left[\boldsymbol{\Xi}_{\bf k,q}\right]^T\cdot\mathcal{H}_{\bf k}\cdot\boldsymbol{\mathcal{V}}_{\bf k}\right]\cdot\bar{\boldsymbol{\epsilon}}+\cdots
    \label{eq_mctsummedafm}
\end{align}
\end{widetext}
where we have used
\begin{align}
    \left[\boldsymbol{\mathcal{G}}^0_{\bf k}\right]^T\cdot\boldsymbol{\mathcal{G}}^0_{\bf k}=1~~~~{\rm and}~~~~\boldsymbol{\mathcal{V}}_{\bf k}^T\cdot\mathcal{H}_{\bf k}\cdot\boldsymbol{\mathcal{V}}_{\bf k}=1
\end{align}

The leading order scattering term in Eq.~\ref{eq_mctsummedafm} is the term under the integral in the equation above. Using explicit calculations gives
\begin{widetext}
\begin{align}
    \bar{\boldsymbol{\epsilon}}^T\cdot\left[\boldsymbol{\mathcal{V}}_{\bf k}^T\cdot\mathcal{H}_{\bf k}\cdot\boldsymbol{\Xi}^2_{\bf k,q}(t_1)\cdot\boldsymbol{\mathcal{G}}^0_{\bf k-q}(t_1)\cdot\boldsymbol{\mathcal{V}}_{\bf k-q}+\left[\boldsymbol{\mathcal{V}}_{\bf k-q}\right]^T\cdot\left[\boldsymbol{\mathcal{G}}^0_{\bf k-q}\right]^T\cdot\left[\boldsymbol{\Xi}^2_{\bf k,q}\right]^T\cdot\mathcal{H}_{\bf k}\cdot\boldsymbol{\mathcal{V}}_{\bf k}\right]\cdot\bar{\boldsymbol{\epsilon}} = 0\ .
\end{align}
\end{widetext}
Hence this, again like the ferrormagnet, reduces to the spin-wave equations and thereby explicitly connects the spin-wave lifetime with the Lyapunov exponent, similar to Eq. \ref{eq_lambdatau}.

\section{Asymptotics of the free decorrelator \label{app:LSW_asymptotics}}
 For the long time asymptotics of the linear spin wave result, we  use the following stationary phase techniques. 
 
 \subsection{Nearest-neighbour FM}
 Isolating the relevant part of the result in the main text, Eq.~\ref{eq_zerodit_22_main},  we consider the asymptotic scaling at large $t$ of
\begin{align}
\int_{BZ} d^d{\bf k}\int_{BZ}d^d{\bf k'}~\cos[(|\gamma[({\bf k})|-|\gamma({\bf k'})|)t+({\bf k+k'})\cdot{\bf r}] 
\label{eq:app_eq_zerodit_22_main}
\end{align}
The asymptotic scaling therefore is contained in
 \begin{equation*}
 I(t)= \int{d^dk  \, e^{i (|\gamma(\vect{k})|+\vect{v} \vect{k}) t }}
 \end{equation*}
   where we defined $\vect{v} = \vect{r}/t$. This   at large $t$ is amenable to a stationary-phase approximation.

In the case of a FM the dispersion relation is given by $\gamma(\vect{k}) = 2J( d -\sum_{i=1}^d \cos(k_i) )$. This is a smooth function of $k$ and has only isolated critical points. Thus, the asymptotic large $t$ scaling is simply $t^{-d/2}$, which gives $t^{-d}$ when squared as quoted in the main text for the scaling of the decorrelator.

We also note that for $v_i > 2 J$, there are no points of stationary phase, thus, by the principle of non-stationary phase, the integral decays  faster than any power-law in $t$, which defines the light-cone velocity $v_{LC}= 2J \sqrt{d}$ along a body-diagonal of the hypercubic lattice, or $v_i = 2J$ along the  direction of the coordinate axes. Notice that the hypercubic appearance of the propagating fronts is evident as the integral fully factorises over $x,y,z$.

 \subsection{Nearest-neighbour AFM}
 The case of the AFM is more complicated. Considering the results in Eqs.~\ref{eq:app_AFM_SKp}-\ref{eq:app_AFM_SKm} we require the asymptotic scaling of
 \begin{align}
     I_0(t) &= \int d^dk \, e^{i \Gamma_k t}\\
     I_1(t) &= \int d^d k \, \sin(\Gamma_k t) \rho(k)\\
     I_2(t) &= \int d^d k \, \sin(\Gamma_k t) \frac{1}{\rho(k)}
 \end{align}
 where $\Gamma(\vect{k}) = 2J \sqrt{d^2-(\sum_i\cos(k_i))^2}$ and $\rho(\vect{k}) = \sqrt{\frac{d-\sum_i \cos(k_i)}{d+\sum_i \cos(k_i)}}$, and we specialised to the case of $\vect{r}=0$.
 We note that by the same argument as for the FM for $|v|>2 J \sqrt{d}$ the integrals decay exponentially, which in contrast to the FM is fully isotropic however.
 
 We begin with the $I_0$ term. This is already not fully trivial. Firstly, the dispersion is not a smooth function. However, evaluating the contribution due to the non-differentiable part at the centre of the Brillouin zone
 \begin{equation}
     \int d^d k \, e^{i |k| t } \sim 1/t^{d}
 \end{equation}
 we find this contribution to be subleading.
 
 In addition, the points of stationary phase are not isolated, but rather form a $d-1$-dimensional manifold defined via $\sum_i \cos(k_i) = 0$. 
 In 1D this does not change things and we have the same scaling as for the FM, e.g. $t^{-1}$ for the decorrelator.
 In higher dimensions this complicates things. Furthermore, while in $2D$ generically the critical points have a non-vanishing Hessian, for special points, e.g. $(k_x,k_y) = (\pi,0)$ the Hessian vanishes.
 
 We consider these contributions in turn. A (d-1) manifold of normal saddles gives a contribution of the form
 \begin{equation}
     \int d^d k \, e^{i (k-k_0)^2 t}  \sim 1/\sqrt{t}
 \end{equation}
 
 In 2D we also have higher order saddles of the form
 \begin{equation}
     \int dk_x dk_y \, e^{i (k_x -\pi)^2 k_y^2 t} \sim \log(t)/\sqrt{t}
 \end{equation}
 
 Thus, in 2D and 3D we obtain additional contributions dominant compared to the normally expected $t^{-d/2}$ scaling, and there is no clear powerlaw behaviour for the AFM in 2D and 3D.
 
 Next we consider the $I_1$ and $I_2$ terms. The additional presence of $\rho(k)$ ($1/\rho(k)$) does not change the asymptotic scaling for those critical points where $\rho(k)$  ($1/\rho(k)$) attains a finite value. Thus, we do not need to revisit the discussion of the $d-1$ manifold where $\sum \cos(k_i) =0$, and only need to consider the effect on isolated critical points.
 
 In 1D we need to additionally consider the points $k=\pi$ ($k=0$), where $\rho$ ($1/\rho$) diverge. For both of these points, expanding in a Taylor series the asymptotic scaling follows from
 \begin{align*}
     \int dk \, \sin( |k| t)/|k| \sim const
 \end{align*}
  
In 2D we need to treat the points $(k_x,k_y)=(0,0)$ and $(k_x,k_y)=(\pi,\pi)$. Again expanding $\rho(k)$ ($1/\rho(k)$) and $\Gamma(k)$ around these points in a polar coordinate system the leading contribution becomes  
\begin{equation}
    \int dk k \, \sin( k t)/k  \sim 1/t
\end{equation}

In 3D the same procedure expanding in spherical coordinates around $(0,0,0)$ and $(\pi,\pi,\pi)$ yields the same integral and a $1/t^2$ contribution.

\subsection{Comparison}
Thus, for the FM we generically obtain powerlaw decay in time, whereas for AFM we obtain a constant contribution to the decorrelator in 1D which will dominate at long times, and complicated crossovers between different scalings in 2D and 3D. 

In addition, the stationary phase calculation suggests a cubic wavefront for the FM and an isotropic one for the AFM as borne out in the explicit calculations and full numerical calculations.

\section{Numerics\label{app:numerics}}
In this section we provide details of the numerical simulations. The set-up parallels our previous work \cite{PhysRevLett.121.250602}, with differences arising due to the broken spin rotational symmetry and finite magnetisation in the ordered regime.

\subsection{Initialisation by Monte Carlo}
We first generate equilibrium spin configurations sampled from the canonical Boltzman distribution at temperature $T$ of the model, Eq.~\ref{eq_ham}, via Monte-Carlo simulations. In the equilibration phase starting from a random spin configuration we perform $10^5$ update sweeps, consisting of 10 micro-canonical overrelaxation lattice sweeps followed by one heatbath sweep. After the equilibration phase we take measurements of the observables, e.g. the full spin configuration. Between each measurement we perform additional updates, consisting of 3 overrelaxation sweeps combined with one heat-bath sweep, either until 100 such combined updates have been performed or until the spin-configuration is decorrelated as set by $\sum_i S^{old}_i  \cdot S_{i}^{new}/N < 0.01$ in the paramagnetic regime, and $\sum_i S^{old}_{\perp,i}  \cdot S_{\perp,i}^{new}/N < 0.01$ in the low temperature ordered regime, where $S_{\perp}$ is the normalised component perpendicular to the order parameter.

\subsection{Integration of the equations of motion}
Starting from these spin configuration we obtain the dynamics by integrating the equations of motion, Eq.~\ref{eq_eom}, using a 8-th order Runge-Kutta solver with fixed time-step. We choose the time-step such that at the final integration time the error of the conserved quantities, here the energy and magnetisation, is below $10^{-6}$, e.g. $|E_f - E_i| < 10^{-6}$.

\subsection{The decorrelator}
To compute the decorrelator, Eq.~\ref{eq_decorrelator}, we evolve each configuration by integrating the equations of motion for the original and the perturbed copy according to the prescription in Eq.~\ref{eq_dels} for a fixed $\epsilon$ simultaneously, or by integrating the equations of motion, Eq.~\ref{eq_eom}, together with the linearised equations for the difference field, Eq.~\ref{eq_deltas}. In the linearised equations of motion the limit $\varepsilon \rightarrow 0$ has been taken, such that the initial condition for the difference field is simply $ \delta{\bf S}_i=({\bf n}\times {\bf S}_{i,0})\delta_{i,0}$. The linearised decorrelator thus carries no factor of $\varepsilon^2$. This needs to be kept in mind when comparing to the expressions explicitly containing these factors. We then average $10^3$ such trajectories to compute the thermal average.

We note that since the equations of motion preserve the magnetisation, the difference between the two copies is limited at low temperature by the ordered moment, as only the transverse component of the spins can decorrelate which has length of order $\sqrt{1-m^2}$. Similarly, in the linearised equations of motion the growth will be in the transverse components as well. Thus, in principle, one may distinguish the longitudinal (parallel to the ordered moment) and the transverse (perpendicular to the ordered moment) components of the decorrelator, which thus acquires a tensorial structure. For the results presented in this work, we have not addressed this additional structure, considering the sum over all components as defined in Eq.~\ref{eq_decorrelator}. This is indeed dominated by the transverse components.

This ties into another advantage of the approach using the linearised equations of motion. While the exponential growth is still limited to the transverse components, it is not limited in size or time, such that we obtain a clean exponential growth up to arbitrary long times, rather than only over a finite window as in the full non-linear equations of motion between two copies.

\subsection{Role of finite magnetisation}
We discuss two additional complication due to the finite magnetisation. Firstly, when comparing to a perturbed copy, one may be concerned that our perturbation, Eq.~\ref{eq_dels}, does not preserve the order parameter. Of course, for small $\varepsilon$ this change is also small. Still, we performed simulations with a modified prescription, where two neighbouring spins are rotated around the order parameter by the angle $\varepsilon$ instead, with no difference in the results.

Secondly, the equations of motion result in a precession of all spins around the magnetisation vector. For the FM the order parameter is the total magnetisation and thus constant in time. However, for the AFM, the order parameter is the staggered moment which is not exactly preserved under the dynamical evolution. While the dynamics of the staggered moment slows down as system size increases, to vanish in the thermodynamic limit, it is present on finite systems. We have therefore computed all observables for the AFM in a static frame, and a dynamic frame defined by a rotation which is chosen such that the staggered moment remains constant. This is primarily important when distinguishing the longitudinal and transverse components which depend on the orientation of the order parameter, rather than the total magnetisation.

\end{appendix}

\bibliography{biblio}

\end{document}